\documentclass[useAMS, usenatbib]{mn2e}
\usepackage{graphicx}
\usepackage{longtable}
\usepackage{multicol}
\usepackage{lscape}
\usepackage{subfig}
\usepackage{natbib}

\title[Q1131+16: a clear view of the inner face of the torus?]{The forbidden high ionisation line region of the type 2 quasar Q1131+16: a clear view of the inner face of the torus?}
\author[M. Rose, C. N. Tadhunter, J. Holt, C. Ramos Almeida and S. P. Littlefair]{M. Rose$^{1}$\thanks{E-mail:m.rose@sheffield.ac.uk (MR); c.tadhunter@sheffield.ac.uk (CNT)}, C. N. Tadhunter$^{1}$\footnotemark[1], J. Holt$^{2}$,  C. Ramos Almeida$^{1}$ and S. P. Littlefair$^{1}$\\
$^{1}$Department of Physics and Astronomy, University of Sheffield, Sheffield S3 7RH\\ 
$^{2}$Leiden Observatory, Leiden University, P.O. Box 9513, 2300 RA Leiden, Netherlands}
\begin{document}

\date{}

\pagerange{\pageref{firstpage}--\pageref{lastpage}} \pubyear{2010}

\maketitle

\label{firstpage}

\begin{abstract}
We present spectroscopic observations of the type 2 quasar SDSS J11311.05+162739.5 (Q1131+16 hereafter; z=0.1732), which has the richest spectrum of forbidden high ionisation lines (FHIL, e.g. [Fe \textsc{vii}], [Fe \textsc{x}], [Fe \textsc{xi}] and [Ne \textsc{v}]) yet reported for an AGN, as well as unusually strong [O \textsc{iii}]$\lambda$4363 emission. The study of this object provides a rare opportunity to investigate the physical conditions and kinematics of the region(s) emitting the FHILs. By comparison with photoionisation model results, we find that the FHIL region has high densities (10$^{5.5}$ $<$ $n_H$ $<$ 10$^{8.0}$ cm$\textsuperscript{-3}$) and ionisation parameters (-1.5 $<$ log[U] $<$ 0), yet its kinematics are similar to those of the low ionisation emission line region detected in the same object (FWHM $\sim$ 360$\pm$30 km/s), with no evidence for a significant shift between the velocity centroid of the FHILs and the rest frame of the host galaxy. The deduced physical conditions lie between those of the Broad-Line (n$_H$$>$10$^9$ cm$\textsuperscript{-3}$) and Narrow-Line Regions ($n_H$$<$10$^6$ cm$\textsuperscript{-3}$) of active galactic nuclei (AGN), and we demonstrate that the FHIL regions must be situated relatively close to the illuminating AGN (0.32 $<$ $r_{FHIL}$ $<$ 50pc). We suggest that the inner torus wall is the most likely location for the FHIL region, and that the unusual strength of the FHILs in this object is due to a specific viewing angle of the far wall of the torus, coupled with a lack of dust on larger scales that might otherwise obscure our view of the torus.

\end{abstract}


\section{Introduction}

Most Seyfert galaxies show some spectral lines from forbidden transitions of highly ionised ions in their spectra, e.g. [Fe \textsc{vii}] , [Fe \textsc{x}], [Fe \textsc{xi}] and even [Fe \textsc{xiv}] \citep{penston}. These emission lines are often blueshifted with respect to the rest frame of the AGN and have velocity widths which are between those of the Narrow-Line Region (NLR), and the Broad-Line Region (BLR, \citealt{penston}, \citealt{appenzeller}, \citealt{mullaney}). In some rare cases many forbidden high ionisation lines (FHILs) of relatively high equivalent width have been detected. Examples include III Zw 77 \citep{osterbrock2}, Tololo 0109-383 \citep{fosbury} and ESO 138 G1 \citep{alloin}. 

The physical mechanisms and conditions that allow strong FHILs to be produced have been debated for some time: whether there is a continuity from the photoionisation processes which produce the lower ionisation lines in the NLR (\citealt{korista}, \citealt{ferg}), or there is an entirely different mechanism for their formation (e.g. collisional excitation in a high temperature gas; \citeauthor{nussbaumer1} \citeyear{nussbaumer1}). 

Given that the transitions associated with many of the FHILs have high critical densities (n$_c$$>$10$^5$ cm$^{-3}$), it has been suggested that they may originate in the innermost torus wall facing the illuminating source. Studies of the high critical density [Fe \textsc{vii}]$\lambda$6086 emission line across Seyfert galaxies from types 1-2, including intermediate types, find that its strength increases relative to the low ionisation lines from type 2 to type 1. This can be interpreted in terms of the orientation-based division between Seyfert types (e.g. \citealt{ant}): as the Seyfert type gets closer to a Seyfert 1, more of the emission from the inner torus becomes visible to the observer (see \citealt{murayama2}, \citealt{nagao1}). Therefore the high critical density lines that are preferentially emitted by the torus are likely to be stronger in type 1 objects, as observed. However, to date the idea that the [Fe \textsc{vii}] emission lines are associated with the torus has not been thoroughly tested using detailed emission line ratios that measure physical conditions accurately.

Although it seems plausible that at least some of the FHIL emission arises in the torus, the causes of the diversity in both the relative strength and kinematics of the FHILs remain uncertain. One possibility is that the unusually strong FHILs observed in some objects are symptomatic of a recent energetic occurrence which is able to illuminate the region that produces these lines \citep{komossa2}; suggestions for such events include the rapid accretion of material from the ISM, the tidal disruption of a stars by the supermassive black hole (SMBH), gamma-ray bursts and supernovae \citep{komossa2}. 

The unusual strength of the FHILs in objects such as III Zw 77, Tololo 0109-383 and ESO 138 G1 provides us with a rare opportunity to study the nature of the region of the AGN which emits them, and thus helps us to better understand the structure of AGN in general.

This paper reports an investigation of the object Q1131+16, which has the richest spectrum of FHILs yet reported for an AGN. Q1131+16 was discovered in the 2MASS survey and was classified as a Seyfert 2 galaxy with a redshift of 0.174. It belongs to sample of quasar-like objects identified on the basis of their red near-IR colours (J-K$>$2.0) in the 2MASS survey \citep{cutri}. In this paper we present deep optical and infrared spectra of Q1131+16, and use these to deduce the physical conditions and kinematics of the FHIL emission lines, with the aim of investigating their origin. The cosmological parameters used throughout this paper are adopted from WMAP: $H_\circ$ = 71 km s$^{-1}$, $\Omega_M$ = 0.27 and $\Omega_\Lambda$ = 0.73 \citep{spergel}, resulting in a spatial scale of 3.46 kpc arcsec$^{-1}$.   

\section{Observations and data reduction}

\subsection{ING ISIS observations}

Low resolution optical spectroscopic observations of Q1131+16 were taken on 9th
February 2007 with the ISIS dual-arm spectrograph on the 4.2-m William Herschel
Telescope (WHT) on La Palma as part of a spectroscopic survey of a representative
sample red, quasar-like objects. The full sample
for this survey comprised a complete RA-limited sub-sample of 24 objects with 
(J-K$>$2) and $z<0.2$ selected from the list of \citealt{hutchings03}, which
is itself representative of the population of red, 2MASS-selected quasars. The
unusually strong FHIL in Q1131+16 were discovered serendipitously in the
course of this survey.

In the red, the R158R grating was used
with the REDPLUS CCD, and in the blue, the R300B grating was used with
the EEV12 CCD. A dichroic at 5300\AA~was employed to obtain spectra with useful
wavelength ranges $\sim$3250-5250\AA~in the blue, and
$\sim$5200-9500\AA~in the red. To reduce the effects of
differential refraction, all exposures were taken when Q1131+16 was at
low airmass (sec $z$ $<$ 1.1) and with the slit aligned close to the
parallactic angle. The seeing for the night of the observations varied from 0.8 to 1.3 arcseconds (FWHM)\footnote{Unfortunately, due to a lack of suitable imaging observations, an accurate seeing estimate does not exist for the exact time of the WHT observations of Q1131+16.}. 

Sets of three 600s exposures were taken on both arms simultaneously, giving a total exposure time
of 1800s. The data were taken with a 1.5 arcsecond slit along a position angle PA315. To 
eliminate contamination from second order emission, a GG495 blocking filter was introduced into the ISIS red arm.

The data were reduced in the standard way (bias subtraction, flat
fielding, cosmic ray removal, wavelength calibration, flux
calibration) using packages in {\sc iraf}\footnote{IRAF is distributed by the National Optical Astronomy Observatory, which is
operated by the Association of Universities for the Research in Astronomy,
Inc., under cooperative agreement with the National Science Foundation (
http://iraf.noao.edu/).}. The
two-dimensional spectra were also corrected for spatial distortions of
the CCD. To reduce wavelength calibration errors due to flexure of the
telescope and instrument, arc spectra were taken at the position of the
object on the sky; the estimated wavelength calibration accuracy is 0.1\AA\ in both the blue and red respectively (however, this may be an underestimate at the extreme edges of the spectra). The atmospheric absorption features were removed by dividing by the spectrum of a telluric standard (BD+17~2352), taken close in time and airmass to the observations of Q1131+16. The spectral resolution, calculated using the widths of
the night sky emission lines, was 6.11 $\pm$ 0.74\AA~in the blue and
10.96 $\pm$ 0.83\AA~in the red (measured in the observed frame of the spectrum). The spatial pixel scales of the 2D spectra are 0.4 arcseconds in the blue and 0.44 arcseconds in the red, and the relative flux calibration uncertainty -- based on 11 observations of 8 flux standard stars
taken throughout the run --  is estimated to be $\pm$5\%.

The spectra were extracted and analysed using the {\sc starlink}
packages {\sc figaro} and {\sc dipso}. 

\subsection{Gemini GMOS observations}

In order to test the possibility that the emission lines of Q1131+16 are variable, a follow-up spectrum was taken using GMOS on the Gemini south telescope at Cerro Pachon.

Spectroscopic and imaging observations using GMOS were taken on the 21st February 2010 during Gemini queue observation time as part of the program GS-2009B-Q-87. The spectral observations comprised three 600 second exposures using a central wavelength 5000\AA\ at an airmass of 1.46, and a 1.5 arcseconds slit aligned along position angle PA163$^{\circ}$  --- close to the parallactic angle for the centre of the observations. This resulted in a spectrum with a useful wavelength range $\sim$3700-6450\AA. The observations of the flux standard star, Feige56, had the same set-up but with an exposure time of 10 seconds, taken at an airmass of 1.61 and with a 5 arcsecond slit. Deep GMOS images were also taken using an r' filter at an airmass of 1.56. Each exposure of the imaging observations had an exposure time of 250 seconds, and the data were taken in a 4 point dither pattern, resulting in a total exposure time of 1000s. The seeing was estimated to be 0.75$\pm$0.02 arcseconds (FWHM), based on measurements of stars in the GMOS image.

The spectra were extracted and analysed using the {\sc starlink}
packages {\sc figaro} and {\sc dipso}, and the {\sc iraf} packages {\sc gemini} and {\sc noao}. Analysis of the night sky lines implies a wavelength calibration uncertainty of 0.50$\pm$0.06\AA\ for the spectrum. The spectral resolution, calculated using the widths of
the night sky emission lines, was 7.12 $\pm$ 0.11\AA\ in the observed frame of the observations, and the spatial pixel scale is 0.147 arcseconds per pixel.

Comparisons between the WHT and Gemini spectra demonstrate that the fluxes of the stronger emission lines (e.g. [O \textsc{iii}]$\lambda\lambda\lambda$4363,4959\&5007, [Ne \textsc{v}]$\lambda$3426, [Fe \textsc{vii}]$\lambda$ 3759) and continuum agree within 20\% between the two sets of observations. This is remarkable considering that the two sets of observations were taken with different instruments, at different slit position angles and under different seeing conditions. 

The GMOS image was reduced using the standard Gemini pipeline reduction software. The photometric calibration used the zero point magnitude for the r' filter advertised on the GMOS south website, taking full account of the airmass and exposure time of the observations. 

\subsection{ING LIRIS observations}

On the night starting the 2nd March 2010 we obtained a 1.388 -- 2.419$\mu$m
spectrum of Q1131+16 using the HK grism in LIRIS \citep{man} on the 4.2-m William Herschel Telescope on La Palma.
The spectrum has a dispersion of 9.7\AA\ per pixel, and a 1 arcsecond
slit was used, giving a spectral resolution of $\sim$40\AA.
The seeing was approximately 2 arcseconds, and the night suffered from
heavy cirrus. A total exposure time of 88 minutes was obtained over
airmasses ranging from 1.0 to 2.2. The total exposure was divided into
240 second sub-exposures and the telescope was nodded in the standard
manner to aid sky subtraction. A random jitter of 10 arcseconds was
added to the default nod positions to limit the impact of bad pixels.
Observations of the A5V star BD+16 2325 were also taken to correct for
the effects of telluric absorption and to provide a relative flux calibration.
Both stars were observed with a slit position angle of zero degrees. For a more detailed description of the reduction process, see \citet{cristina}.

\section{Results}

\begin{center}
\begin{table*}
\caption{Centroid positions and spatial extents of selected emission lines and the continuum at different wavelength intervals, as measured from the 2D WHT spectrum. The centroids (measured in pixels) and the FWHM (measured in arc seconds) were determined using single Gaussian fits to the bright cores of the spatial distributions of the emission lines along the slit.}
\begin{tabular}{cccccc}
\hline

$\lambda_{Rest}$ \AA & Range \AA & Centroid & $\pm$ & FWHM (arc seconds) & $\pm$\\

\hline
\it{Blue Arm}\\
\hline
$[$Ne V] 3425	&	35	&	98.34	&	0.01	&	1.35	&	0.01\\
cont. 3392	&	35	&	98.23	&	0.04	&	1.71	&	0.04\\
$[$O II] 3727	&	45	&	98.39	&	0.08	&	1.56	&	0.06\\
cont. 3690	&	45	&	98.21	&	0.07	&	1.78	&	0.06\\
$[$Fe VII] 3759	&	45	&	98.32	&	0.07	&	1.28	&	0.05\\
$[$Ne III] 3868	&	45	&	98.32	&	0.01	&	1.24	&	0.01\\
cont. 3923	&	45	&	98.33	&	0.06	&	1.81	&	0.07\\
H$\delta$	&	50	&	98.49	&	0.05	&	1.04	&	0.05\\
cont. 4147	&	50	&	98.25	&	0.06	&	1.74	&	0.06\\
H$\gamma$	&	35	&	98.33	&	0.04	&	1.09	&	0.04\\
$[$O III] 4363	&	35	&	98.35	&	0.05	&	1.14	&	0.02\\
cont. 4466	&	35	&	98.38	&	0.05	&	1.73	&	0.07\\
\hline
\it{Red Arm}\\
\hline
He II 4686	&	50	&	89.82	&	0.12	&	1.25	&	0.15\\
cont. 5545	&	50	&	89.84	&	0.04	&	1.74	&	0.05\\
H$\beta$	&	50	&	89.74	&	0.04	&	1.09	&	0.04\\
$[$O III] 4959\&5007	&	130	&	89.75 	&	0.01	&	1.18	&	0.01\\
cont. 5093	&	130	&	89.85	&	0.04	&	1.74	&	0.05\\
$[$Fe VII] 6086	&	65	&	89.83	&	0.02	&	1.19	&	0.02\\
cont. 6139	&	65	&	89.80	&	0.04	&	1.87	&	0.05\\
H$\alpha$	&	145	&	89.81	&	0.01	&	1.09	&	0.02\\
cont. 6435	&	145	&	89.81	&	0.04	&	1.71	&	0.05\\
$[$Fe XI] 7892	&	40	&	89.76	&	0.10	&	1.18	&	0.11\\
cont. 7927	&	40	&	89.84	&	0.06	&	1.80	&	0.07\\

\hline 
\end{tabular}
\end{table*}
\end{center}

\begin{figure*}
\centering
\includegraphics[scale=0.3]{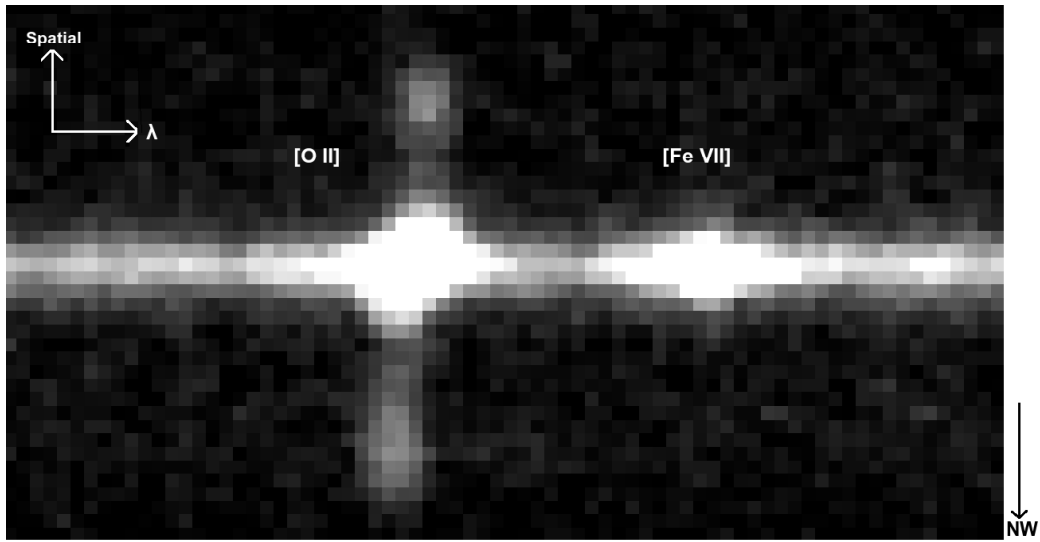}
\caption{A section of the 2D WHT spectrum along PA315 showing [O \textsc{ii}]$\lambda$3727 and [Fe \textsc{vii}]$\lambda$3759. Note that the [O \textsc{ii}] is extended 6.9 arcseconds above and 8.1 arseconds below, the continuum of Q1131+16 and is well resolved. The spatial direction of the slit is indicated along side the plot. In addition, a nuclear velocity gradient in the [O \textsc{ii}] can be seen from the fact that its profile is slanted when compared to that of [Fe \textsc{vii}]}.
\end{figure*}

\subsection{Spatial distribution of the emission lines}

While our 2D spectra show that all the emission lines are strongly concentrated on the nucleus of the host galaxy, we detect spatially extended line emission in [O \textsc{ii}]$\lambda$3727 (see Figure 1), H$\beta$, [O \textsc{iii}]$\lambda\lambda$5007,4959 and the H$\alpha$+[N \textsc{ii}] blend. The emission lines have a maximum extent of 9 arcseconds in the north--south direction (29kpc) in the Gemini spectrum (PA163), and 15 arcseconds in the north west--south east direction (47kpc) in the WHT spectra (PA315), corresponding to the spatial extent of the galaxy disk visible in our Gemini r' image (see $\oint$3.5).

In order to investigate the nuclear spectrum, we extracted apertures of size of 1.5x1.5 arcseconds from the 2D WHT and Gemini spectra. The apertures were centred on the nucleus.

To determine the spatial distributions of both the emission lines and continuum along the slit, spatial slices were extracted from the 2D WHT spectrum over the wavelength ranges given in Table 1. The continuum slices were extracted with similar wavelength ranges to their nearby emission lines so that the continuum could be accurately subtracted from the slices containing the emission lines. The {\sc dipso starlink} package was then used to measure the centroid and spatial FWHM of the flux distribution of each slice, by fitting a single Gaussian profile to the central cores of the emission.

Overall, the centroids measured for each emission line and continuum slice in Table 1 are all consistent within their uncertainties, implying no spatial offset between any of the emission lines individually, nor between the emission lines and the continuum. 

The spatial FWHM of all the emission lines --- with the exception of [O \textsc{ii}]$\lambda$3727 --- are consistent with the seeing on the night of the observations (see $\oint$2.1), suggesting that their spatial distributions are unresolved  in the observations. Therefore the spatial distributions of the high ionisation lines provide an indication of the true seeing of the WHT observations (1.1 $<$ FWHM $<$ 1.35 arcseconds).
As expected, there is evidence that the seeing degrades towards shorter wavelengths, since the
two shortest wavelength FHIL have significantly larger spatial FWHM than their longer wavelength counterparts.  Moreover, fact that the spatial FWHM measured for all continuum slices significantly exceeds the estimated seeing, demonstrates that the continuum emission is spatially resolved in the nuclear regions.

The spatial FWHM of the [O \textsc{ii}]$\lambda$3727 emission line is significantly broader than the other FHILs at similar wavelengths in Table 1, and this line is clearly resolved in the 2D spectrum (see Figure 1). The combination of the large spatial extent and spatially resolved nuclear emission of [O \textsc{ii}] makes this emission line ideal for the study of the emission line kinematics across the host galaxy Q1131+16 presented in $\oint$3.6.

\subsection{Line identifications}

The optical spectra of Q1131+16 are shown in Figures 2-5. What makes the spectrum of Q1131+16 special is not only the large number of emission lines, but also their rich variety. In particular, the detected FHILs include [Fe \textsc{v}]$\lambda\lambda\lambda$3839,3891\&4181, [Fe \textsc{vi}]$\lambda\lambda\lambda$5146,5176\&5335, [Fe \textsc{vii}]$\lambda\lambda\lambda\lambda\lambda\lambda$3759,4893,5159,5276,5720\&6086, [Fe \textsc{x}]$\lambda$6375, [Fe \textsc{xi}]$\lambda$7891 and [Ne \textsc{v}]$\lambda\lambda$3346\&3426\footnote{In this study we define a FHIL as an emission species with an ionisation potential greater than or equal to 54.4 eV (that of He \textsc{ii}).}. As well as the FHILs, lower ionisation species which are not typically detected in quasar spectra are found in the spectrum, including [Fe \textsc{iv}]$\lambda\lambda\lambda\lambda$2829,2836,4903\&5236, along with Bowen resonance fluorescence lines of O \textsc{iii}, such as O \textsc{iii} $\lambda$3133. Also detected are more typical AGN emission lines such as H$\alpha$, H$\beta$, He \textsc{ii} $\lambda$4686, [O \textsc{ii}]$\lambda\lambda$3726,3729, [O \textsc{iii}] $\lambda\lambda\lambda$4363,4959\&5007. However, the FHILs are unusually strong with respect to these latter lines, with [Ne \textsc{v}] $\lambda$3426 comfortably exceeding the strength of the [O \textsc{ii}]$\lambda$3727 doublet, and [Fe \textsc{vii}]$\lambda$6086 of comparable strength to H$\beta$. Also notable is the unusual strength of [O \textsc{iii}]$\lambda$4363 compared to H$\gamma$ and [O \textsc{iii}]$\lambda$5007 ([O \textsc{iii}]5007/4363 = 5.86$\pm$0.21 in the WHT spectrum), as well as the relatively modest ratio of [O \textsc{iii}]$\lambda$5007 to H$\beta$ ([O \textsc{iii}]/H$\beta$=5.27$\pm$0.16), given the strength of the other high ionisation lines.

The strongest emission lines in this object have relatively large equivalent widths (EWs), including a large number of FHILs with EWs comparable to other typical lower ionisation emission lines. Indeed, the EWs of the FHILs in Q1131+16 are larger than those of any other AGN with published spectra, including other objects with unusually strong FHILs such as III Zw 77 \citep{osterbrock2} and Tololo 0109-383 \citep{fosbury}. Moreover the [O \textsc{iii}] emission line luminosity of Q1131+16 (4.6$\pm$0.1x10$^{8}$ L$_\odot$) would lead to its classification as a quasar 2 object according to the criterion of \citet{zakamska}. Overall, the rich variety of emission lines, and their relatively large EWs, allow for a thorough investigation of the physical conditions of the FHIL emission region.

\begin{figure*}
\centering
\includegraphics[scale=0.55, angle=270]{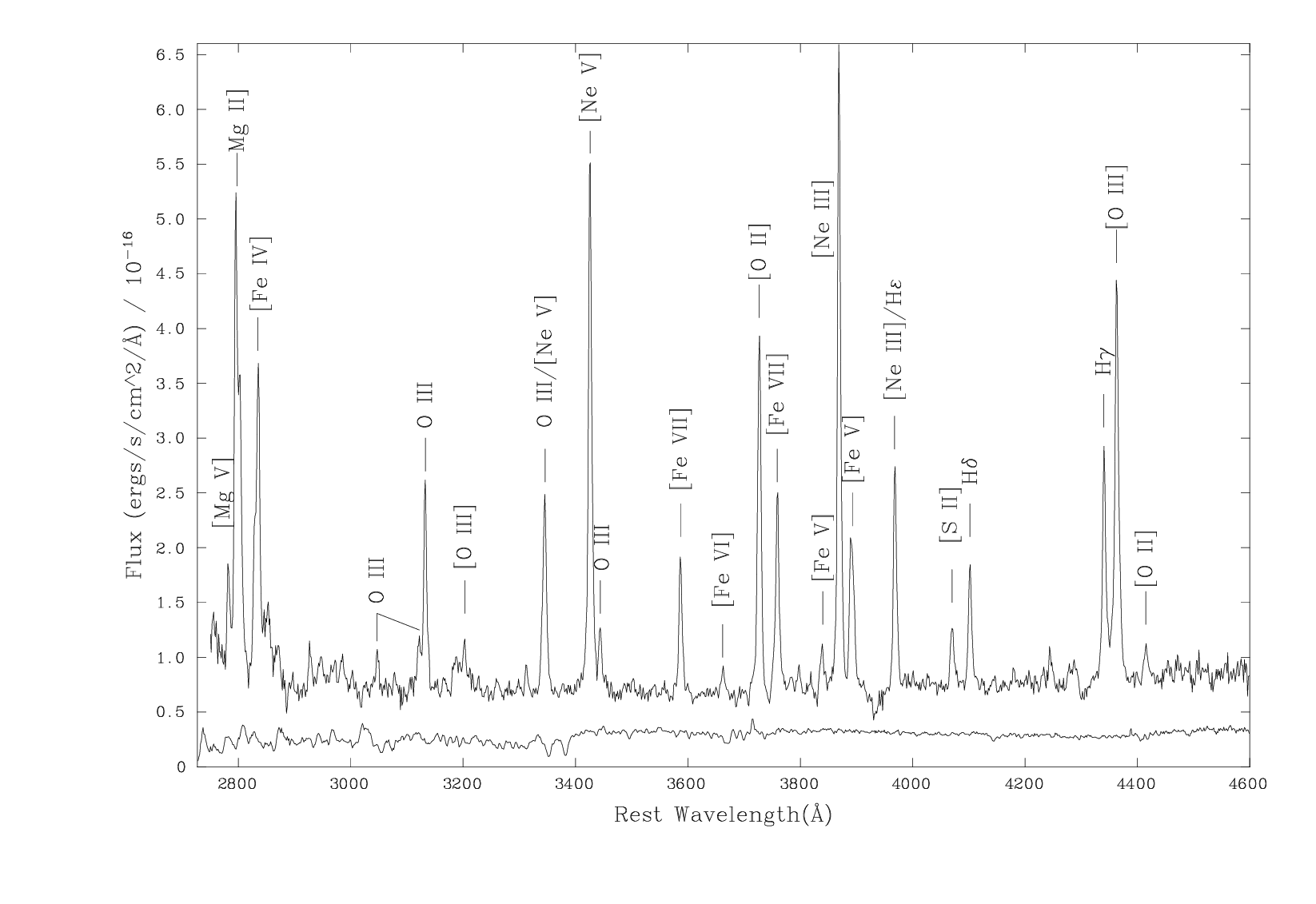}
\caption{WHT nuclear spectrum of Q1131+16 taken on the ISIS blue arm. Double lines which are labeled with an '\&' indicates that both components are resolvable, those lableled with a '/' indicates that both components are not resolvable. The flux scale is measured in units of 10$^{-16}$ ergs s$^{-1}$ \AA$^{-1}$ cm$^{-2}$. Note the strength of high ionisation lines such as [Ne \textsc{v}] and [Fe \textsc{vii}] relative to [O \textsc{ii}], and of [O \textsc{iii}]$\lambda$4363 relative to H$\delta$. For reference we also show a scaled version of the night sky spectrum extracted from the 2D frames at the bottom of the plot.}
\includegraphics[scale=0.55, angle=270]{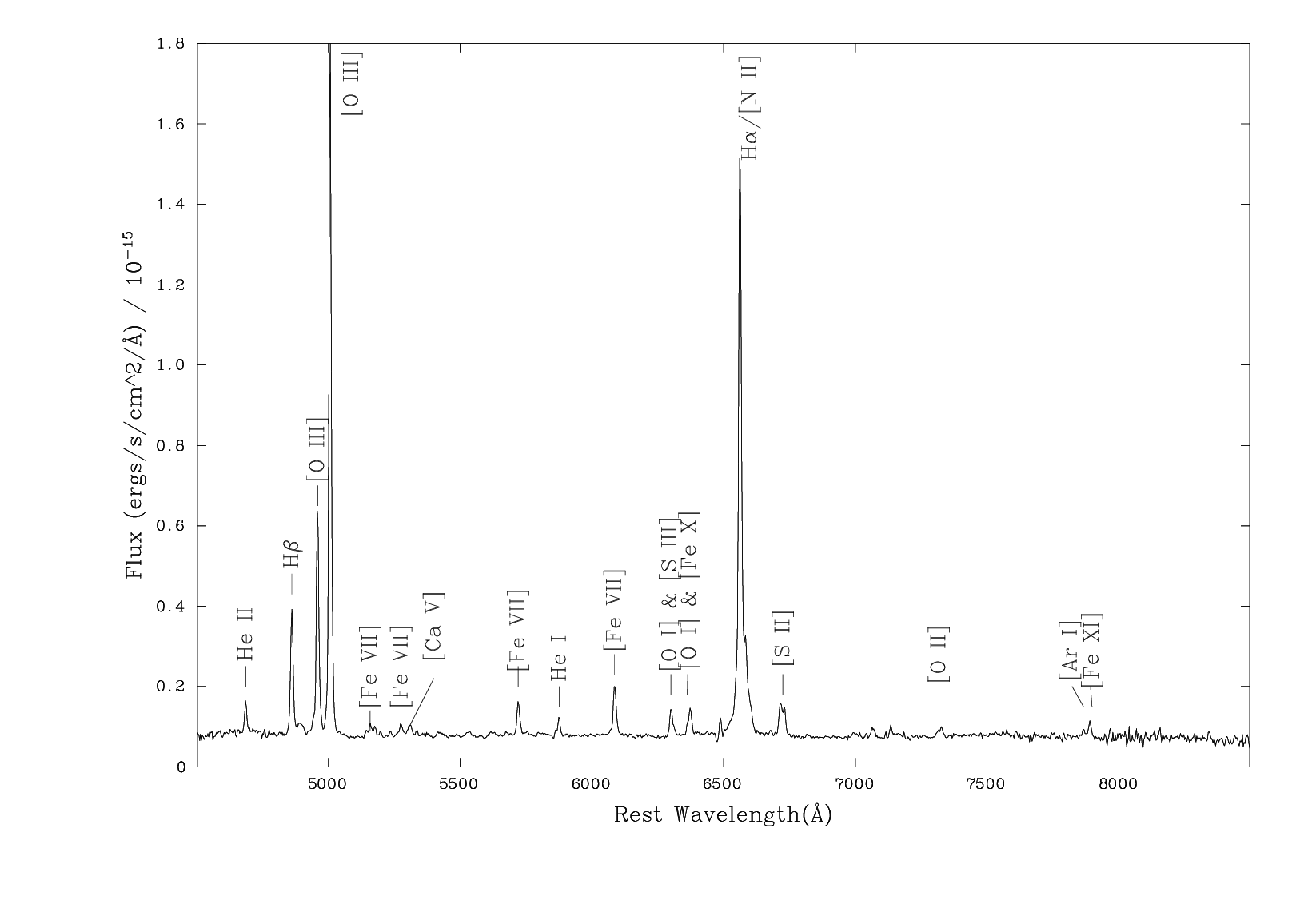}
\caption{WHT nuclear spectrum of Q1131+16 taken on the ISIS red arm. Double lines which are labeled with an '\&' indicates that both components are resolvable, those lableled with a '/' indicates that both components are not resolvable. The flux scale is measured in units of 10$^{-15}$ ergs s$^{-1}$ \AA$^{-1}$ cm$^{-2}$.}
\end{figure*}

\begin{figure*}
\centering
\includegraphics[scale=0.55, angle=270]{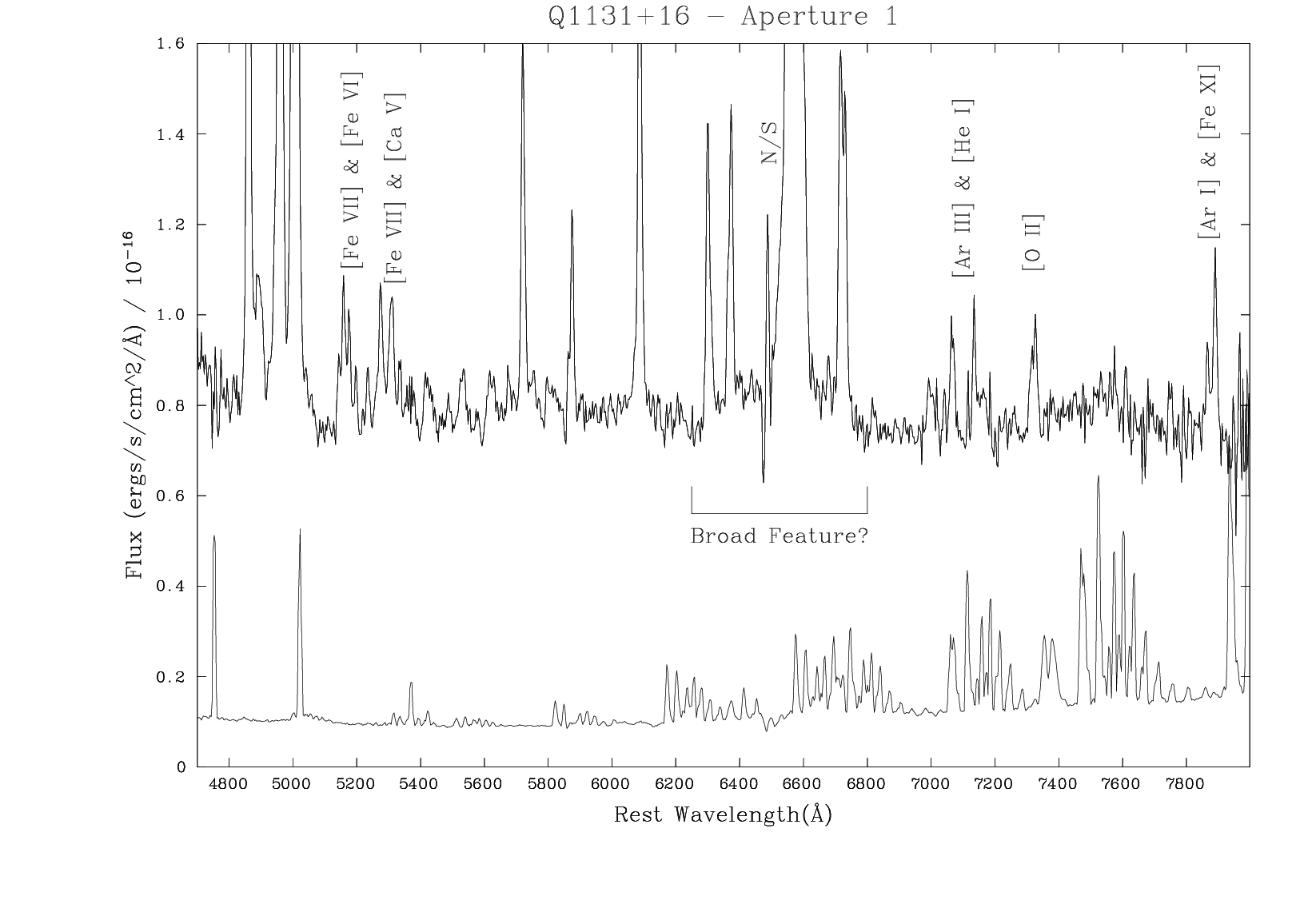}
\caption{An expanded plot of the Q1131+16 ISIS red arm spectrum. This has been presented to highlight the weaker features in the red spectrum. The flux scale is measured in units of 10$^{-16}$ ergs s$^{-1}$ \AA$^{-1}$ cm$^{-2}$. As well as the multitude of high ionisation  lines, note the broad base to the H$\alpha$$+$[N \textsc{ii}] blend. For reference we also show a scaled version of the night sky spectrum extracted from the 2D frames at the bottom of the plot.}
\includegraphics[scale=0.55, angle=270]{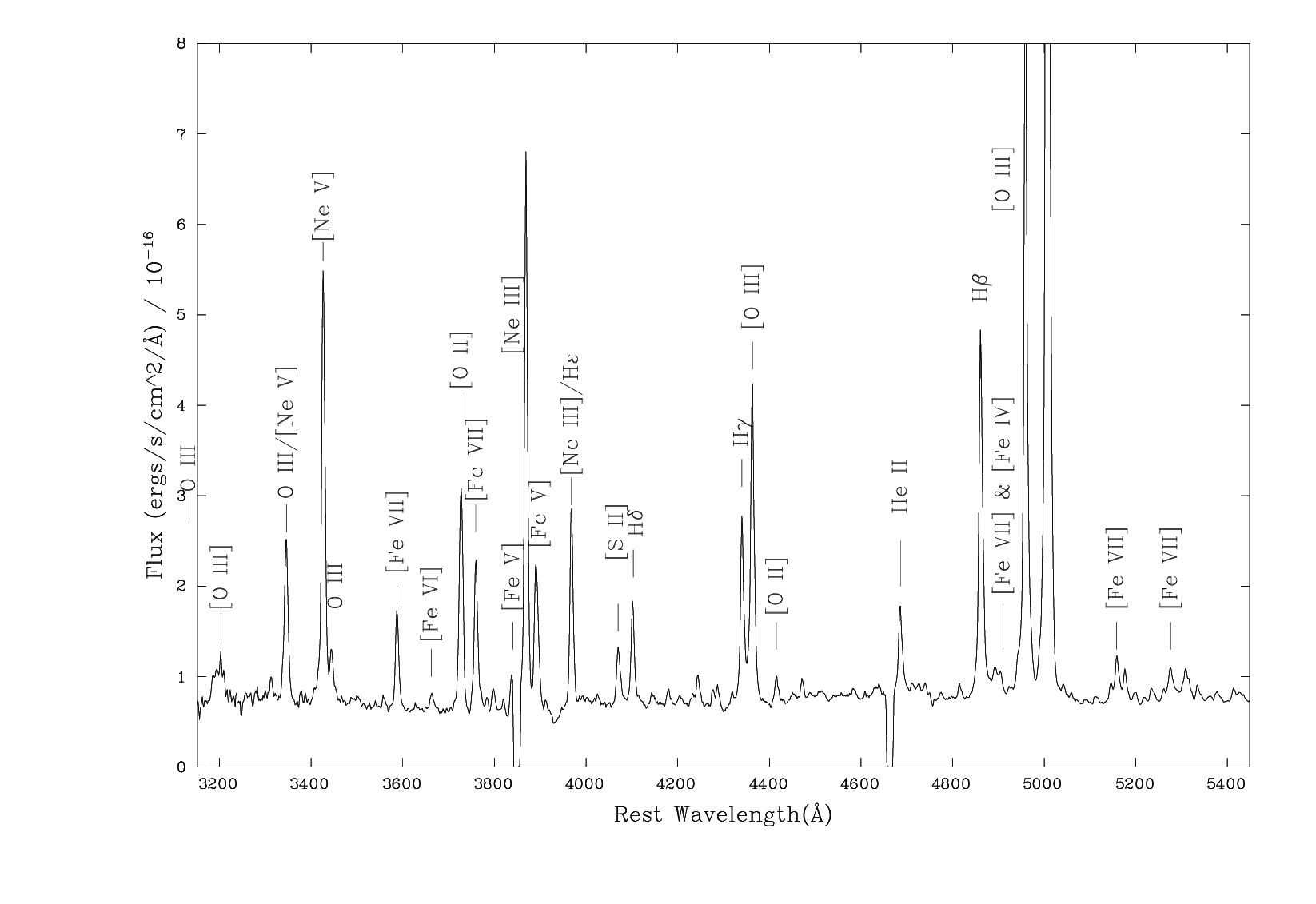}
\caption{ Nuclear spectrum of Q1131+16 taken using GMOS on the Gemini South telescope. The flux scale has been narrowed here to highlight how rich the spectrum is in emission lines. The flux scale is measured in units of 10$^{-16}$ ergs s$^{-1}$ \AA$^{-1}$ cm$^{-2}$. The gaps in the spectrum are due to the gaps between the CCD chips in the GMOS instrument.}
\end{figure*}

A full list of line identifications made from these spectra is presented in Table 5\footnote{Unfortunately tables 5 \& 6 are not available here, however they are available in the source tar.}. All the line identifications have been determined by fitting single Gaussians to the emission features in both the WHT and Gemini spectra. The rest-frame wavelength ranges that were fitted for Q1131+16 are 2700-4600 \AA\ and 4400-8000 \AA\ in the WHT spectra, and 3100-5450 \AA\ in the Gemini spectrum. Table 5 gives the line flux ratios relative to H$\beta$ for both spectra. The fluxes of the emission lines have not been corrected for intrinsic reddening for reasons that will become clear in $\oint$3.8. In addition, it appears that there is no need of a Galactic extinction correction, since the IRSA extinction tool in the NED gives a reddening of only E(B-V)= 0.0306 \citep{schlegel}.

\begin{figure*}
\includegraphics[scale=0.55]{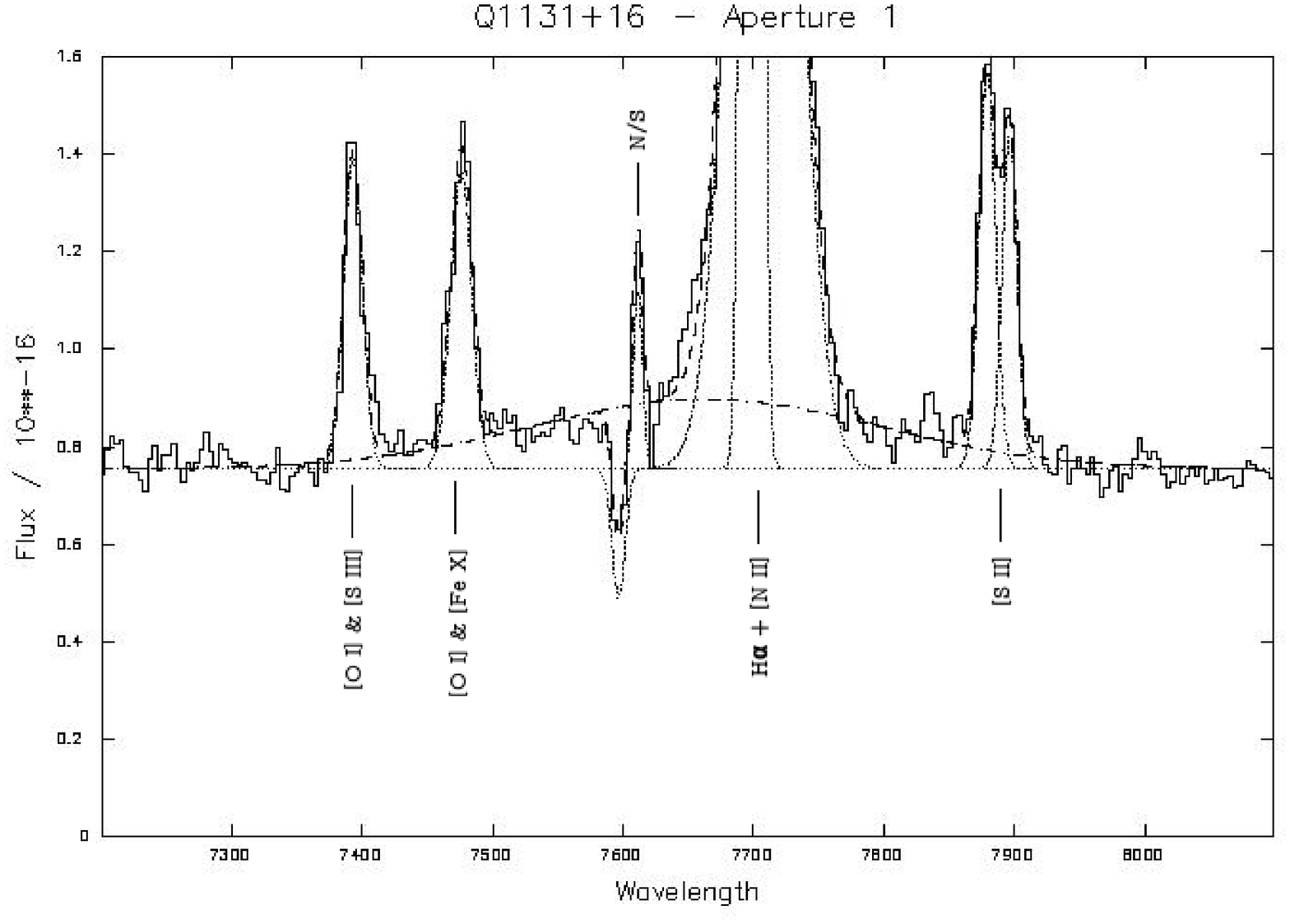}
\caption{The broad component centered on the H$\alpha$$+$[N \textsc{ii}] emission blend. The telluric feature is fitted as well as the various emission components from Q1131+16.}
\end{figure*}

The majority of line identifications have been confirmed in other astrophysical objects such as AGN and planetary nebula (e.g. \citealt{osterbrock2}, \citealt{fosbury}, \citealt{alloin}, \citealt{kaler}). However, there are several emission lines on Table 5 which have been identified using the National Institute of Standards and Technology (NIST) spectral line database (see ref. 4 in Table 5). An emission line was only regarded as a secure ID if its line centre was within 1.5 sigma of the wavelength
predicted for that particular emission line based on the mean redshift, and if its S/N ratio exceeded 3.0. As well as the 66 identified emission lines, there are 37 emission lines which remain unidentified (see Table 6), in the sense that there are no IDs for them in the NIST spectral line database that give redshifts within 1.5 sigma of the mean redshift, yet their S/N exceeds 3. Note that, due to its higher S/N, the Gemini spectrum reveals a large number of faint lines that were not detected in the original WHT spectrum (see Tables 5 and 6).

Interestingly, although there are many iron FHILs in the spectrum of Q1131+16, there is no clear evidence for the [Fe \textsc{xiv}]$\lambda$5303 emission line. The identification of this emission feature at $\sim$5300 \AA\ has been controversial in the past: it has been debated whether it is [Fe \textsc{xiv}]$\lambda$5303 or [Ca \textsc{v}]$\lambda$5309 (e.g. \citealt{oke}, \citealt{weedman}). When this feature is fitted in our WHT spectrum, if only the [Ca \textsc{v}]$\lambda$5309 identification is considered, the individual redshift of the emission line (0.17319$\pm$0.00011 for the WHT spectrum) agrees within the uncertainties with the average redshift of all the lines (0.17325$\pm$0.00001, see $\oint$3.6). However, if the feature is identified with [Fe \textsc{xiv}]$\lambda$5303, the individual redshift becomes 0.17455$\pm$0.00011, which is significantly higher ($>$10$\sigma$) than the mean redshift.

A further interesting feature of the spectrum is that the H$\alpha$$+$[N \textsc{ii}] emission blend shows tentative evidence for a broad base of rest-frame width 11,500$\pm$2200 km s$^{-1}$ (FWHM, see Figures 4 \& 6). This is consistent with the presence of a scattered \citep{am}, or directly observed, broad-line region (BLR) component. Spectropolarimetry observations will be required to confirm the scattered BLR possibility.

\begin{figure*}
\centering
\includegraphics[scale=0.55, angle=270]{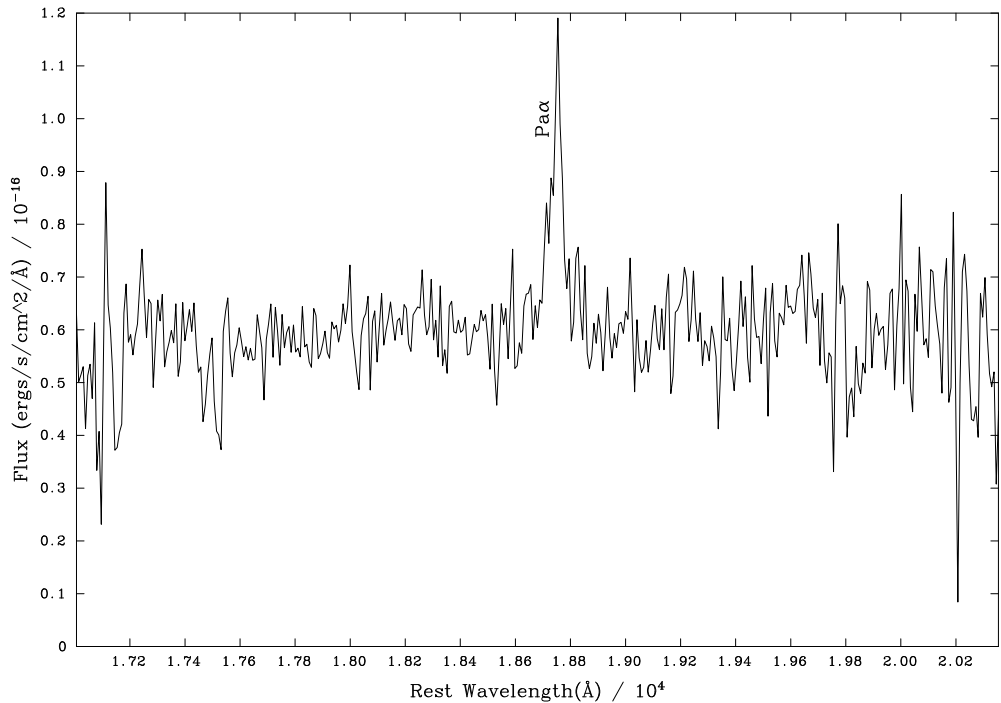}
\caption{The K-band LIRIS spectrum of Q1131+16. The rest wavelength range shown here is 17000-20500 \AA. The only significant feature is the Pa$\alpha$ emission line.}
\end{figure*}

\begin{figure*}
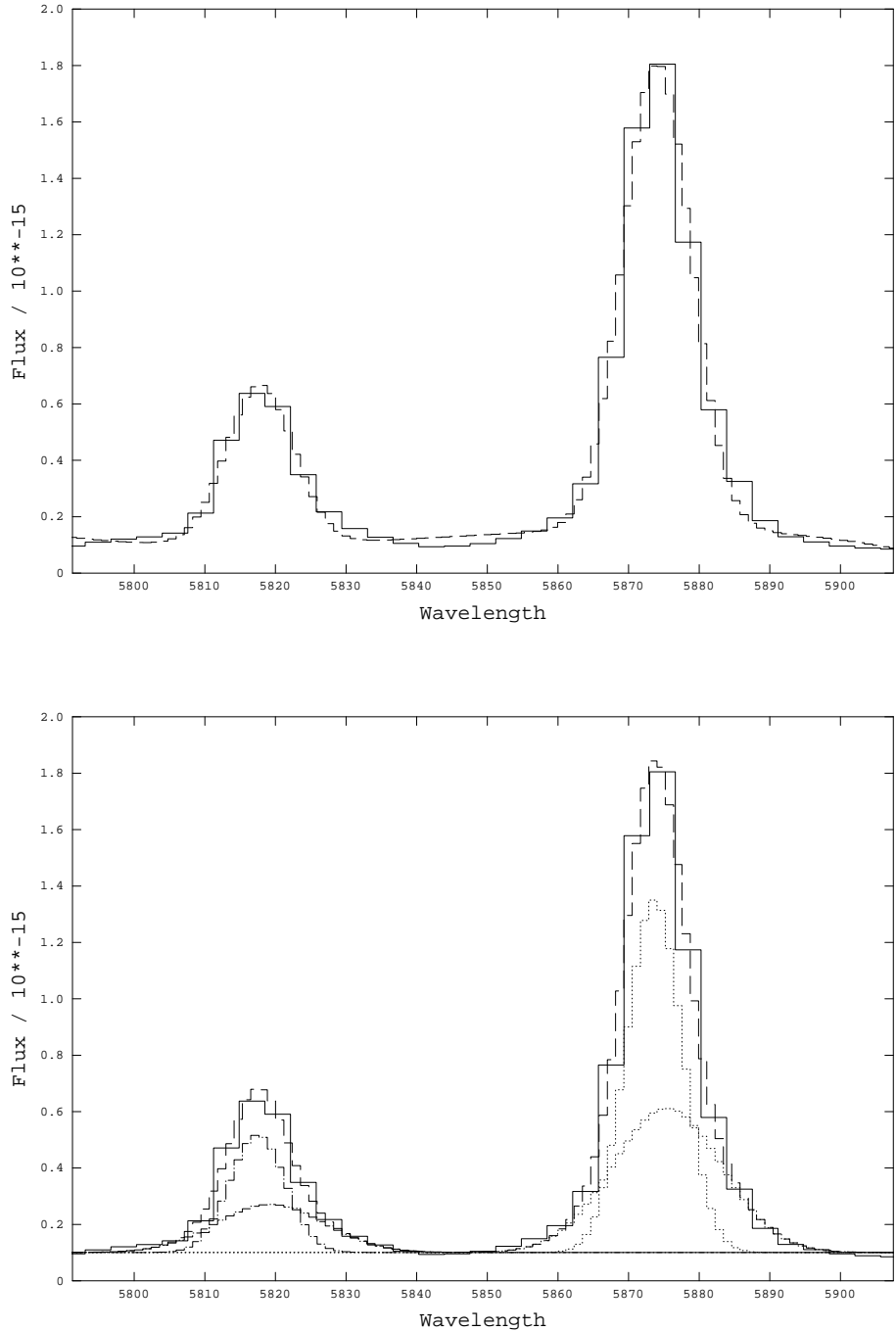

 \includegraphics[scale=0.5, angle=270]{5007singfit.eps}
\includegraphics[scale=0.5, angle=270]{5007modfit.eps}
 \caption{Fits to the [O \textsc{iii}]$\lambda\lambda$5007,4959 emission lines.{\it Top}. Single Gaussian fit to both [O \textsc{iii}]$\lambda$5007 and $\lambda$4959. The overall fit does not fit the wings of the emission lines well. {\it Bottom}. The double Gaussian fit to the [O \textsc{iii}]$\lambda$5007 and $\lambda$4959 emission lines from the Gemini data. Here the model has had more success in fitting the wings of the emission lines when compared to that above. The individual components of the fit are included on the plot; the components of $\lambda$5007 are drawn with a dotted line and the components of $\lambda$4959 are fitted with a dashed and dotted line. The overall line profile is fit with a dashed line. The model throughout this investigation is based on the [O \textsc{iii}]$\lambda$5007 emission line. The flux is measured in 10$^{-15}$ ergs s$^{-1}$ \AA$^{-1}$ cm$^{-2}$ and the wavelength is in units of \AA.}
\end{figure*}

\subsection{Infrared spectrum}

In order to further understand the nature of this object we made spectroscopic observations of Q1131+16 at near-IR wavelengths. The observed wavelength range (1.4 to 2.4$\mu$m) was selected in order to simultaneously detect both Pa$\alpha$ and Pa$\beta$. 

The K-band NIR spectrum of Q1131+16 taken using LIRIS is presented in Figure 7. Due to its relatively low S/N, this spectrum does not show the abundance of emission lines seen at optical wavelengths. Although weak narrow Pa$\alpha$ and Pa$\beta$\footnote{The Pa$\beta$ emission line is on the blue edge of the H-band NIR spectrum, because of this any measured emission line data is questionable.} emission lines are detected, there is no sign of any broad Pa$\alpha$ and Pa$\beta$ components, implying the BLR is enshrouded by dust. A single Gaussian fit to the Pa$\alpha$ line shows that it is unresolved within the uncertainties for the resolution of our observations, and its redshift (z=0.17314$\pm$0.00011) is consistent with those of the optical emission lines.

 Any quasar nucleus component present in this system must be highly extinguished; the relatively red near-IR colours of this source measured by 2MASS do not appear to be due to a moderately extinguished quasar component that becomes visible at the longer near-IR wavelengths. Therefore this object cannot truly be described as a $\lq$red quasar'.

\subsection{Spectral fitting model}

The emission lines in the spectrum of Q1131+16 were initially fitted with single Gaussian profiles, but such fits did not provide an entirely adequate fit to the wings of the stronger lines (see Figure 8). To  overcome this, a double Gaussian model was fitted to the spectral lines. This model is based on the fit of the [O \textsc{iii}]$\lambda\lambda$5007,4959 lines, because they are particularly strong emission features. To produce the model, both broad and narrow Gaussians were fitted to the Gemini [O \textsc{iii}] line profiles, where the centres, widths and intensities of each Gaussian were free parameters. The widths (FWHM) of the narrow components for the [O \textsc{iii}] emission lines were found to be consistent with the instrumental width of the Gemini spectra (7.1 \AA) at 6.9$\pm0.2$ \AA. We therefore use the instrumental width to fit the narrow components in the model fit to other lines, varying this to take into account the resolution of the different spectra. The measured FWHM of the broad component (14.1$\pm$0.6 \AA (FWHM), corresponding to a rest-frame velocity width of 720$\pm$30 km s$^{-1}$) was used to obtain an intrinsic velocity width for the broader component by correcting its FWHM in quadrature using the instrumental width. In addition, the broad component is redshifted by 1.8$\pm$0.1 \AA\ from the narrow component, corresponding to a velocity shift of 92$\pm$6 km s$^{-1}$ in the galaxy rest frame. The double Gaussian [O \textsc{iii}] model was then fitted to all other emission lines, in order to derive the line fluxes listed in Tables 4 and 5.  

Figure 8 shows the spectral fit to the [O \textsc{iii}]$\lambda$5007 emission line used to determine the spectral fitting model. The overall fit in the bottom panel of Figure 8 is significantly better than that shown in the top panel, where the emission lines are fitted with a single component free fit. The parameters from the double Gaussian [O \textsc{iii}] fits were used to fit the emission lines throught the spectrum. In general these fits were extremely successful. There is, however, a minority of lines --- indicated by notes 1 and 2 in Tables 4 and 5 --- which are not fitted well by both components of the [O \textsc{iii}] model. These lines are generally weak, and many are in blends with other emission lines.

The [N \textsc{ii}]$\lambda$$\lambda$6548,6584 doublet is blended with the H$\alpha$ emission line and therefore was modelled using constraints provided by atomic physics (i.e. same FWHM, 1:3 intensity ratio, and line centre of $\lambda$6548 fixed to the centre of $\lambda$6583). This is true of other doublet blends, for example the [O \textsc{i}]$\lambda\lambda$6300,6364 (blended with [S \textsc{iii}]$\lambda$6312 and [Fe \textsc{x}]$\lambda$6375).

\subsection{Host galaxy morphology}

\begin{figure*}
\centering
\includegraphics[scale=0.35, angle=270]{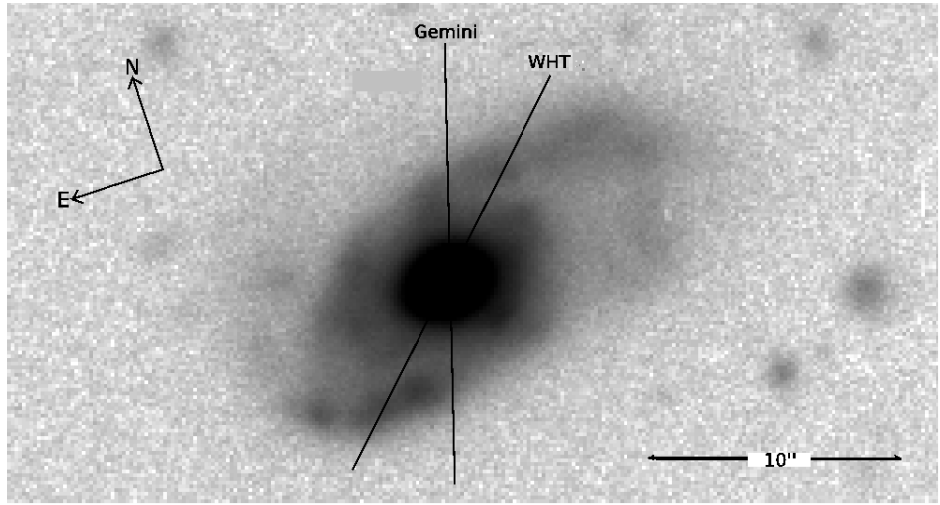}
\caption{GMOS r' image of the host galaxy Q1131+16. Indicated are the North and East axes as well as the slit PA of each spectroscopic data set.}
\end{figure*}

Figure 9 presents our Gemini r'-band image of Q1131+16. The galaxy morphology appears consistent with a late type spiral classification (Sb or Sc), with a moderate degree of asymmetry in the sense that the spiral arms are more extended to the west than the east of the nucleus. However, we do not find evidence for the ``large scale and clear tidal disturbance'' reported by \citet{hutchings03}. Clearly, a more detailed analysis of the morphology and kinematics of the host galaxy is required in order to determine the true interaction status of this system. 

Based on aperture photometry, the total magnitude of the host galaxy was estimated using a circular aperture of diameter 23 arcseconds (57 kpc), giving an r'-band magnitude of 17.11$\pm$0.26 --- consistent with the Petrosian r-band magnitude for this object listed on the SDSS web site. Note that this estimate represents an upper limit on the brightness of the stellar component of the host galaxy since no correction has been made for AGN components such as emission lines and scattered light.

\subsection{Emission line kinematics and redshift}

The median redshift determined using all of the securely identified emission lines is z=0.17325$\pm$0.00001. From only the securely identified FHILs (IP $\geq$ 54.4 eV) the redshift is z=0.17323$\pm$0.00002, while for the low ionisation lines we find z=0.17326$\pm$0.00001. These values are consistent within their uncertainties, as emphasised by Figure 10, where estimated redshifts of all the lines are plotted. 

In order to check whether the mean emission line redshift represents that of the host, the [O \textsc{ii}]$\lambda$3727 emission line blend was fitted as a function of position across the galaxy for both Gemini and WHT spectra. Figure 11 presents the results. The velocities were determined relative to the median redshift of all the emission lines of the system (z=0.17325$\pm$0.00001), assuming a rest-frame wavelength of 3727.4 \AA\ for the blend --- representing the flux weighted wavelength of the [O \textsc{ii}]$\lambda\lambda$3726,3729 blend for a density of 10$^3$ cm$^{-3}$. For both position angles, the central region ($\pm$1.2 arcseconds, $\pm$4kpc) shows a velocity field that resembles a rotation curve; the velocity gradient of the [O \textsc{ii}]$\lambda$3727 emission in the nucleus can also clearly be seen in Figure 1. This component of emission may correspond to an inner disk of the galaxy, implying that the [O \textsc{ii}]$\lambda$3727 emission component in Q1131+16 is resolved in the nucleus. However, the two datasets differ significantly at larger radii; such differences are likely due to the different slit PAs, given that each slit intersects different regions of Q1131+16 see Figure 11): whereas the slit for the Gemini data is within $\sim$ 20 degress of the minor axis of the host galaxy, that for the WHT is offset for the minor axis by $\sim$ 50 degress. 

The rotation curve dereived from the PA315 WHT data appears flat at large radii. However, the velocities of the flat parts of the velocity curves are asymmetric about the host redshift measured from the nuclear spectrum: the rotation velocity ranges from +90$\pm$5 km/s for the region moving away from the observer to -176$\pm$5 km/s for the region moving towards the observer in the WHT spectrum. This asymmetry can be explained by either the uncertainty in the electron density assumed in determining the rest wavelength of the [O \textsc{ii}] blend ($n_e$ is likely to differ between the nucleus and the disk of Q1131+16), or the uncertainty in the wavelength calibration of the spectrum. Both of these aspects can systematically shift the curve in Figure 11, leading to an asymmetric appearence. However, despite the apparent asymmetry, the bisector of the two flat parts of the rotation curve is within 50 km s$^{-1}$ of the rest frame of the host galaxy estimated on the basis of the median redshift derived from the nuclear spectrum. Thus our data provide no evidence for high velocity outflows in the nuclear emission line regions.

On the other hand, the rotation curve derived from the PA163 Gemini data appears more complex: outside the inner rapidly rotating region, the velocities drop significantly at intermediate radii (1.2$<$r$<$2.5 arcseconds) on either side of the nucleus, before rising again at larger radii. It is possible that the inner rapid rotation along PA163 samples the seeing disk of the same inner disk component detected in the WHT data along PA315. In this case, the initial drop in the velocities (1.2$<$r$<$2.5 arcseconds) may be due to the slit sampling the outer disk close to the minor axis where the (projected) rotation velocity amplitude is likely to be small. However, given the complexity of the rotation curve along PA163, non-circular gas motions cannot be ruled out.

Our double Gaussian [O \textsc{iii}] model fits the majority of the lines well in the nuclear spectrum, suggesting that the velocity widths of the FHIL and low ionisation emission lines are similar.  The median rest-frame velocity width of the emission lines, based on single Gaussian fits, is found to be 361$\pm$29 km s$^{-1}$ (FWHM).

Figures 12 and 13 also show that there is no correlation between the widths of the emission lines as measured by single Gaussian fits, and their critical densities and ionisation potentials. Again this suggests the kinematics of the FHILs and the low ionisation lines are similar. 

\subsection{FHILs in $\lq$typical' Seyfert galaxies}

FHILs are detected in the spectra of all types of Seyfert galaxies \citep{penston}. These emission lines have been studied rigorously in the past (e.g. \citealt{penston}, \citealt{nagao2}, \citealt{mullaney2}), and several common properties have been identified. It is therefore important to highlight the differences between the spectrum of Q1131+16 and the spectra of \lq typical' Seyfert galaxies.

Possibly the most notable property of the FHILs in typical AGN is that the velocity widths (FWHM as measured by single Gaussian fits) tend to be intermediate between the NLR and BLR (in the range of 500 $<$ FWHM $<$ 1000 km s$^{-1}$, \citealt{ao}). However the FHILs in Q1131+16 and other similar objects (e.g. III Zw 77 and Tololo 0109-383 etc.) do not share this property. In the case of Q1131+16, Figures 12 and 13 show no significant variation in the velocity width (FWHM) with the critical density, and ionisation potential.

For typical AGN it is also found that FHILs are blueshifted with respect to the host galaxy rest-frames. In fact, the blueshift correlates with the observed velocity FWHM \citep{mullaney}. There is no evidence for this in the spectrum of Q1131$+$16. The individual redshifts from single Gaussian fits of each emission line are plotted in Figure 10. From this it is clear that there is no tendency for the FHILs to be blueshifted with respect to the average redshift of the low ionisation lines; the points representing the FHIL redshifts are scattered about the average redshift of the low ionisation species. 

Finally, the FHILs of Q1131+16 and similar objects have larger equivalent widths when compared to those of other \lq typical\rq\ Seyfert galaxies.

\begin{figure}
\centering
\includegraphics[scale=0.38]{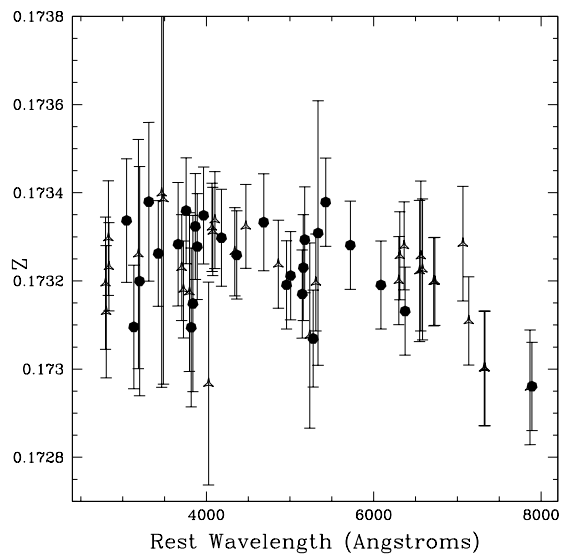}
\caption{Individual redshifts of the emission lines of Q1131$+$16 as measured for the WHT data. The triangles indicate the low ionisation emission lines and neutral species emission, the black circles indicate the FHILs. All points lie within 1.5 $\sigma$ of the mean redshift, however the longer wavelength emission lines ([O \textsc{ii}]$\lambda\lambda$7320\&7330, [Ar \textsc{iv}]$\lambda$7868 and [Fe \textsc{xi}]$\lambda$ 7892) do not. This is likely due to larger wavelength calibration errors at the long wavelength end of the spectrum.}
\end{figure}

\begin{figure}
\centering
\includegraphics[scale=0.38]{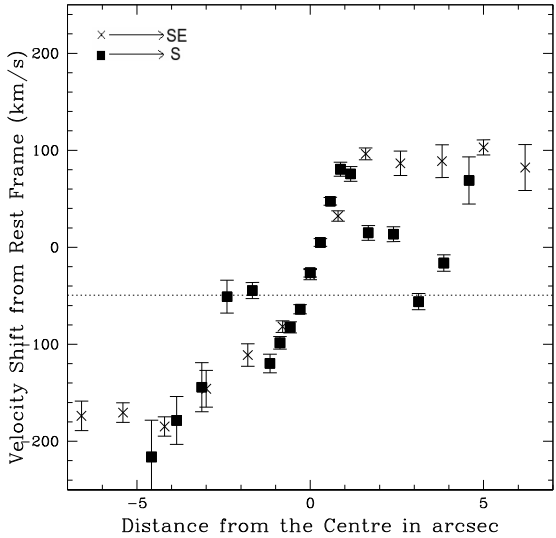}
\caption{The rotation curves for Q1131$+$16. The crosses (WHT, PA315) and filled squares (Gemini, PA163) are the measured velocity shifts determined at different spatial locations along the slit of the long slit spectrum. The directions of the slits are indicated in the upper left-hand corner. Velocity shifts are determined by the difference between the measured redshift of [O \textsc{ii}] on each pixel and the average redshift of the emission lines. The horizontal line indicates the bisector of the flat parts of the rotation curves at large radii along PA315.}
\end{figure}

\begin{figure}
\centering
\includegraphics[scale=0.38]{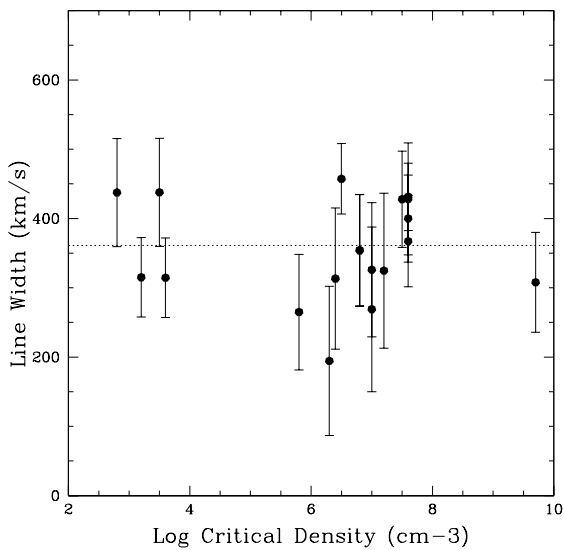}
\caption{Rest-frame line width (FWHM) versus critical density for the emission lines of Q1131+16. The line widths have been determined from single Gaussian free fits to the emission lines, and have been corrected for the instrumental profile.}
\end{figure}

\begin{figure}
\centering
\includegraphics[scale=0.38]{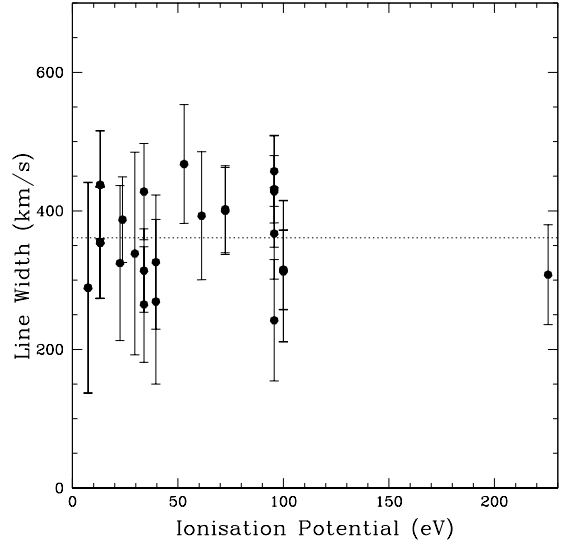}
\caption{Rest-frame line width (FWHM) versus ionisation potential for the emission lines of Q1131+16. The line widths have been determined from single Gaussian free fits to the emission lines, and have been corrected for the instrumental width.}
\end{figure}

\subsection{Balmer decrements}

In order to derive accurate physical conditions, it is crucial to correct for any intrinsic dust extinction. The most common way to achieve this for optical spectra is to use Balmer line recombination ratios. 

Table 2 presents flux ratios for the Balmer recombination lines measured from the Gemini and WHT spectra. The H$\gamma$/H$\beta$ and H$\delta$/H$\beta$ ratios in both spectra agree with their theoretical Case B values within the uncertainties, assuming conditions typical of the NLR. This suggests that there is no need for an extinction correction for the spectrum of Q1131+16. Note that, because of the large equivalent widths of the Balmer lines, uncertainties due to underlying Balmer absorption lines from young stellar populations in the host galaxy are not a serious issue for Q1131+16. 

\begin{table}
\caption{Line flux ratios for Balmer recombination emission lines for both the WHT and Gemini observations. Ratios are relative to H$\beta$. The last three columns on the table are the expected ratios for case B recombination at different temperatures at a typical NLR density (10$^{4}$ cm$^{-3}$, \citealt{osterbrock1}).}
\begin{tabular}{c c c c c}
\hline
Species & WHT & Gemini & 10000K  & 20000K\\ \hline
H$\alpha$ & 4.99$\pm$0.17 & - & 2.85 & 2.74\\
H$\gamma$ & 0.44$\pm$0.02 & 0.45$\pm$0.03 & 0.47 & 0.48\\
H$\delta$ & 0.24$\pm$0.02 & 0.24$\pm$0.01 & 0.26 & 0.26\\
\hline
\end{tabular}
\end{table}

Another interesting feature of the spectrum of Q1131+16 is that its H$\alpha$ flux is high when compared to other Balmer lines in the series. The H$\alpha$/H$\beta$ ratio is not consistent with Case B recombination (H$\alpha$/H$\beta$$=$4.99$\pm$0.17), see $\oint$ 4.2 for a detailed discussion. Given the potential for collisional excitation of the H$\alpha$ line (see $\oint$ 4.2) \citep{gaskell}, we ignore the H$\alpha$/H$\beta$ ratio when assessing the level of extinction in Q1131+16. Therefore, since the high order Balmer line ratios for this source provide no evidence for significant reddening, we make no reddening correction of the line ratios.

\section{Discussion}

\subsection{Physical conditions implied by the FHILs} 

\begin{table}
\centering
\caption{Diagnostic line intensity ratios measured for the nuclear regions Q1131+16, as derived from the WHT data. Values in brackets are the ratios given by the Gemini data. Potentially, the [Fe \textsc{vii}]$\lambda$6086 line may suffer from some contamination by [Ca \textsc{v}]$\lambda$6087. However, the fact that the [Fe \textsc{vii}] (5720/6086) ratio is within 1 $\sigma$ of the value predicted by atomic physics ([Fe \textsc{vii} (5720/6086) = 0.617: \citealt{nussbaumer2}) indicates that any such contamination is modest.}
\begin{tabular}{l l l}
\hline
Species & Ratio & $\pm$\\
\hline 
$[$O \textsc{ii}](3726+3729)/(7317+7330) & 5.02 & 1.19\\
$[$O \textsc{iii}](5007/4363) & 5.86(6.84) & 0.21(0.32)\\
$[$O \textsc{iii}](5007/H$\beta$) & 5.27(5.74) & 0.16(0.16)\\
$[$S \textsc{ii}](6717+6731)/(4069+4076) & 7.51 & 1.46\\
$[$Fe \textsc{vi}](5176)/[Fe \textsc{v}](3891) & 0.267(0.176) & 0.076(0.008)\\
$[$Fe \textsc{vii}](6086/3759) & 1.18 & 0.076\\
$[$Fe \textsc{vii}](5159/6086) & 0.208 & 0.067\\
$[$Fe \textsc{vii}](5720/6086) & 0.667 & 0.067\\
$[$Fe \textsc{x}](6375)/[Fe \textsc{vii}](6086) & 0.473 & 0.074\\
$[$Fe \textsc{xi}](7892)/[Fe \textsc{vii}](6086) & 0.317 & 0.046\\
$[$Ne \textsc{v}](3426)/[Fe \textsc{vii}](6086) & 2.04 & 0.207\\
\hline
\end{tabular}
\end{table}

\begin{figure}
\centering
\includegraphics[scale=0.38]{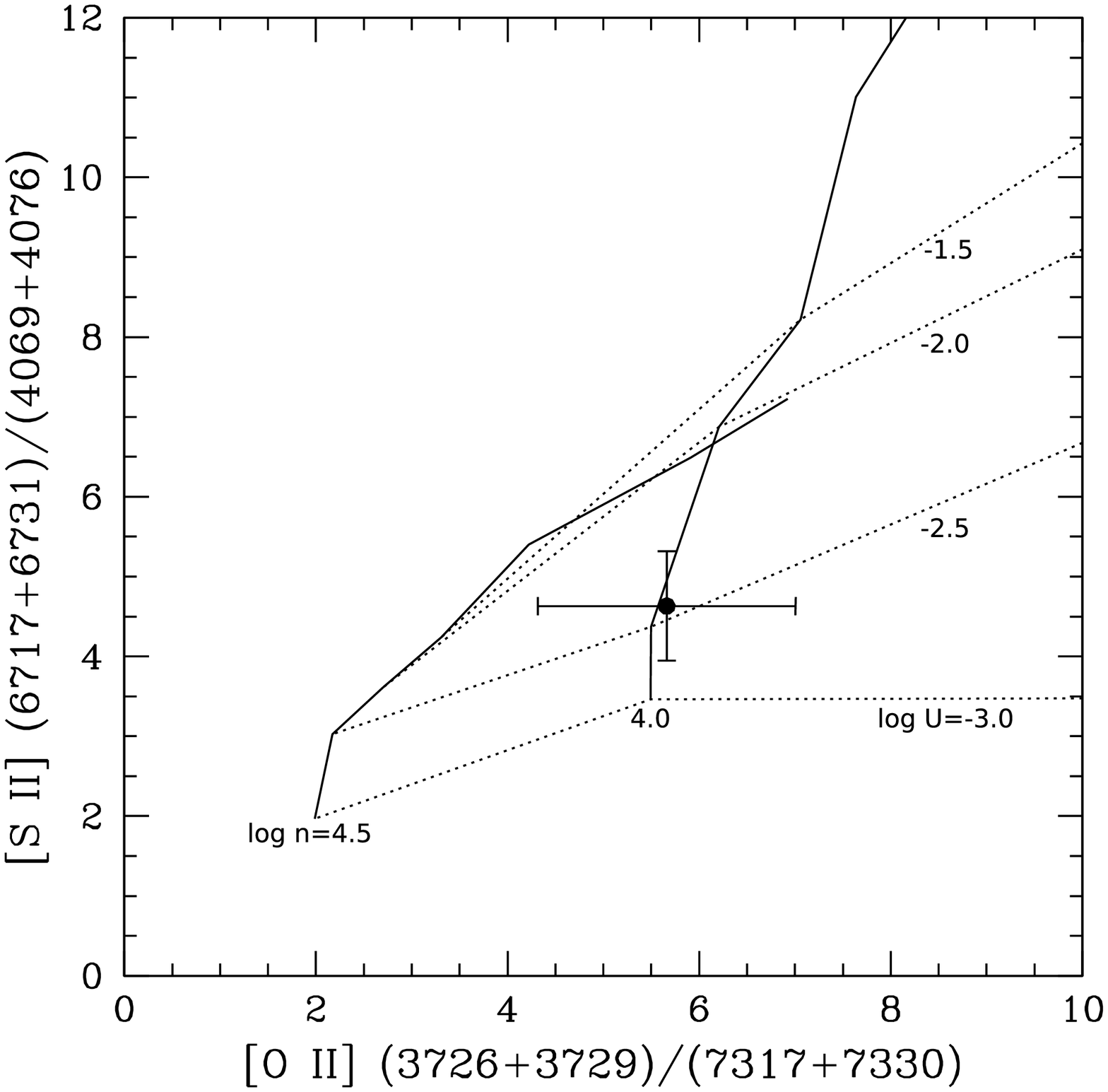}
\caption{Diagnostic plot showing the measured [S\textsc{ii}](6717+6731)/(4069+4076) and [O\textsc{ii}](3726+3729)/(7317+7330) transauroral ratios (filled point), compared with photoionisation model results for different densities and ionisation parameters. Equal densities are connected by the solid lines (as labelled), and the ionisation parameters are connected by the dotted lines (also labelled).}
\end{figure}

\begin{figure}
\centering
\includegraphics[scale=0.38]{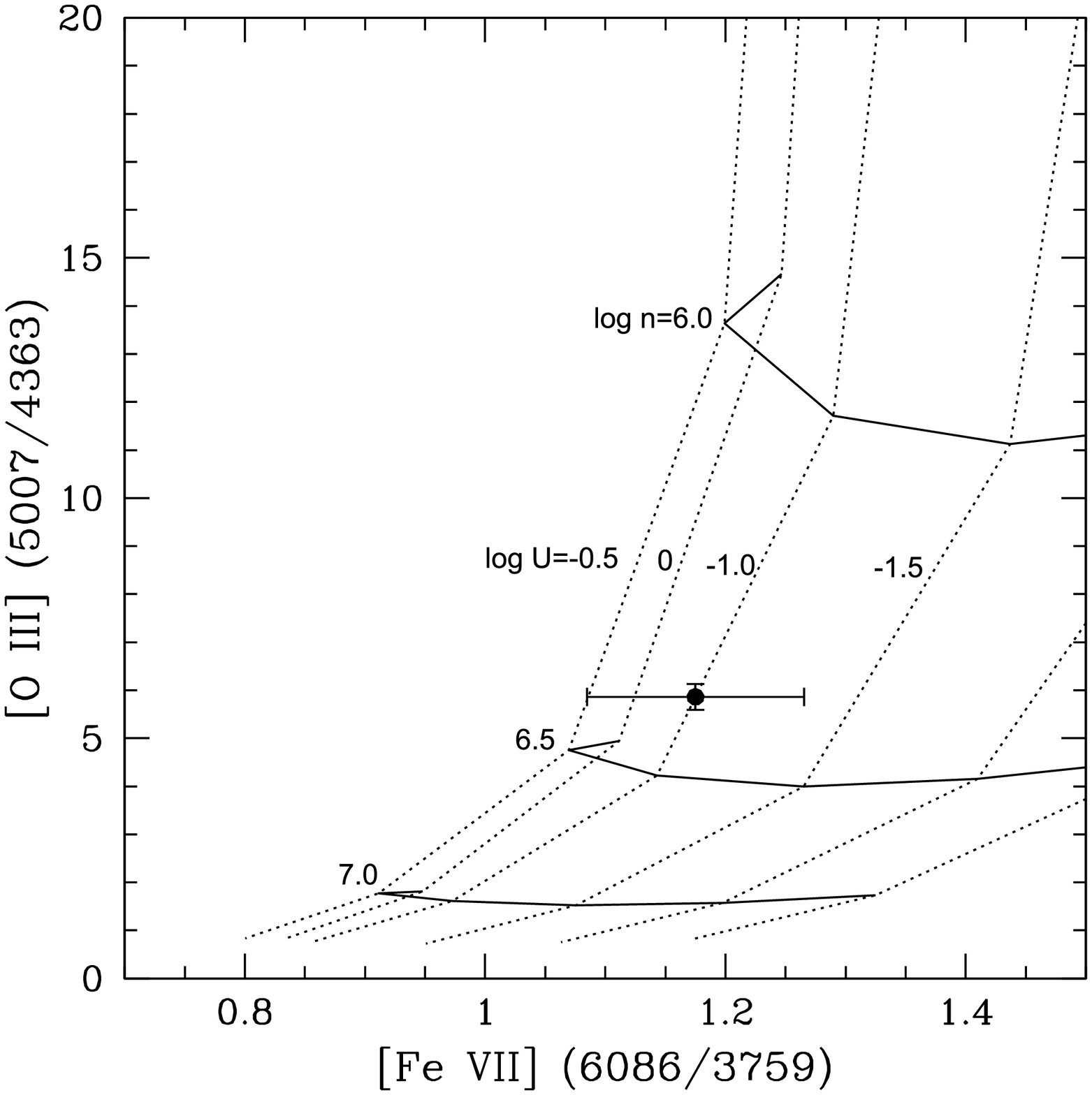}
\caption{Diagnostic plot showing the measured [O \textsc{iii}](5007/4363) and [Fe \textsc{vii}](6086/3759) ratios (filled point), compared with photoionisation model results for different densities and ionisation parameters. Equal densities are connected by the solid lines (as labelled), and the ionisation parameters are connected by the dotted lines (also labelled).}
\end{figure}

\begin{figure}
\centering
\includegraphics[scale=0.38]{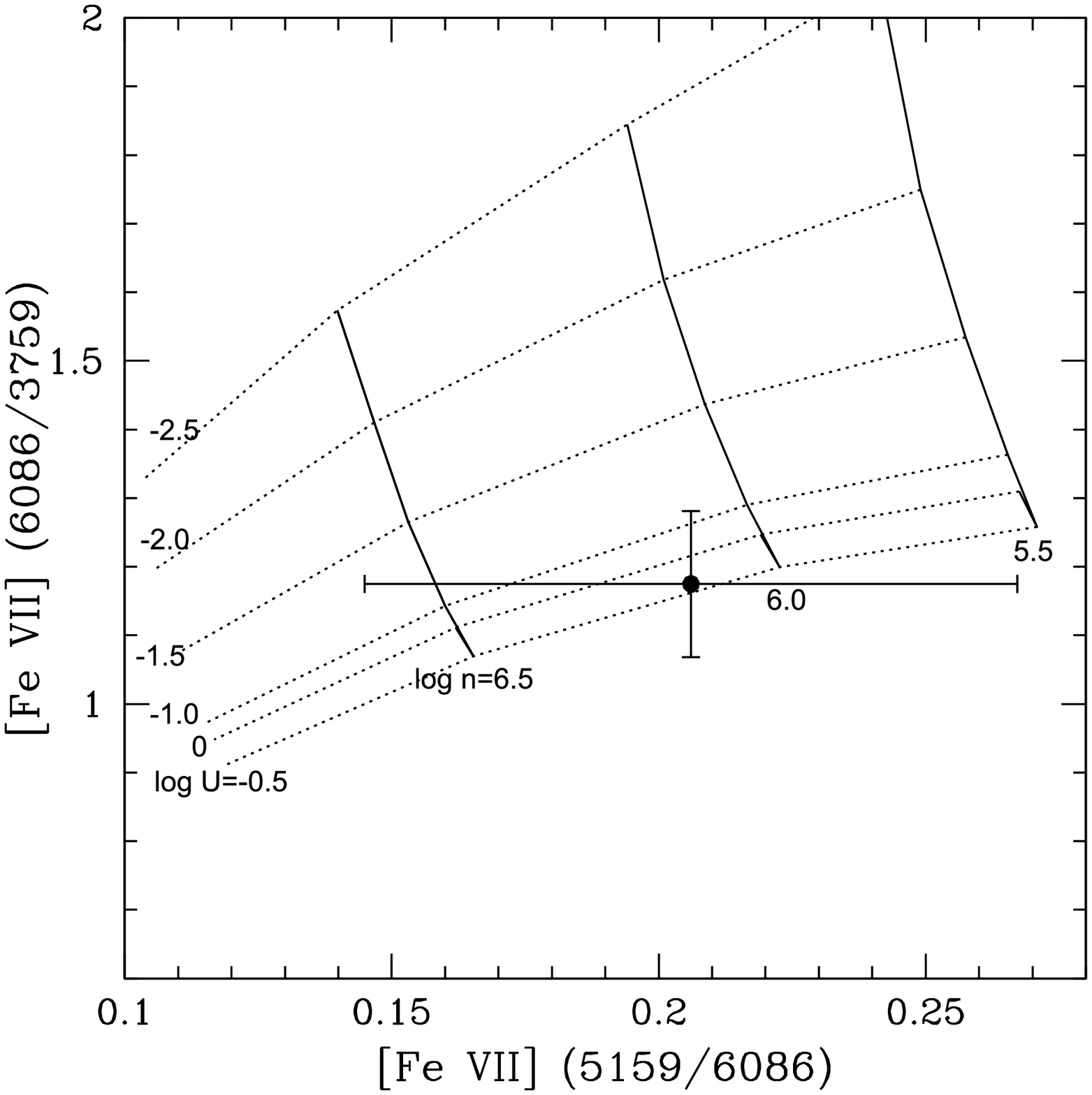}
\caption{Diagnostic plot showing the measured [Fe \textsc{vii}](6086/3759) and [Fe \textsc{vii}](5159/6086) ratios (filled point), compared with photoionisation model results for different densities and ionisation parameters. Equal densities are connected by the solid lines (as labelled), and the ionisation parameters are connected by the dotted lines (also labelled).}
\end{figure}

\begin{figure}
\centering
\includegraphics[scale=0.38]{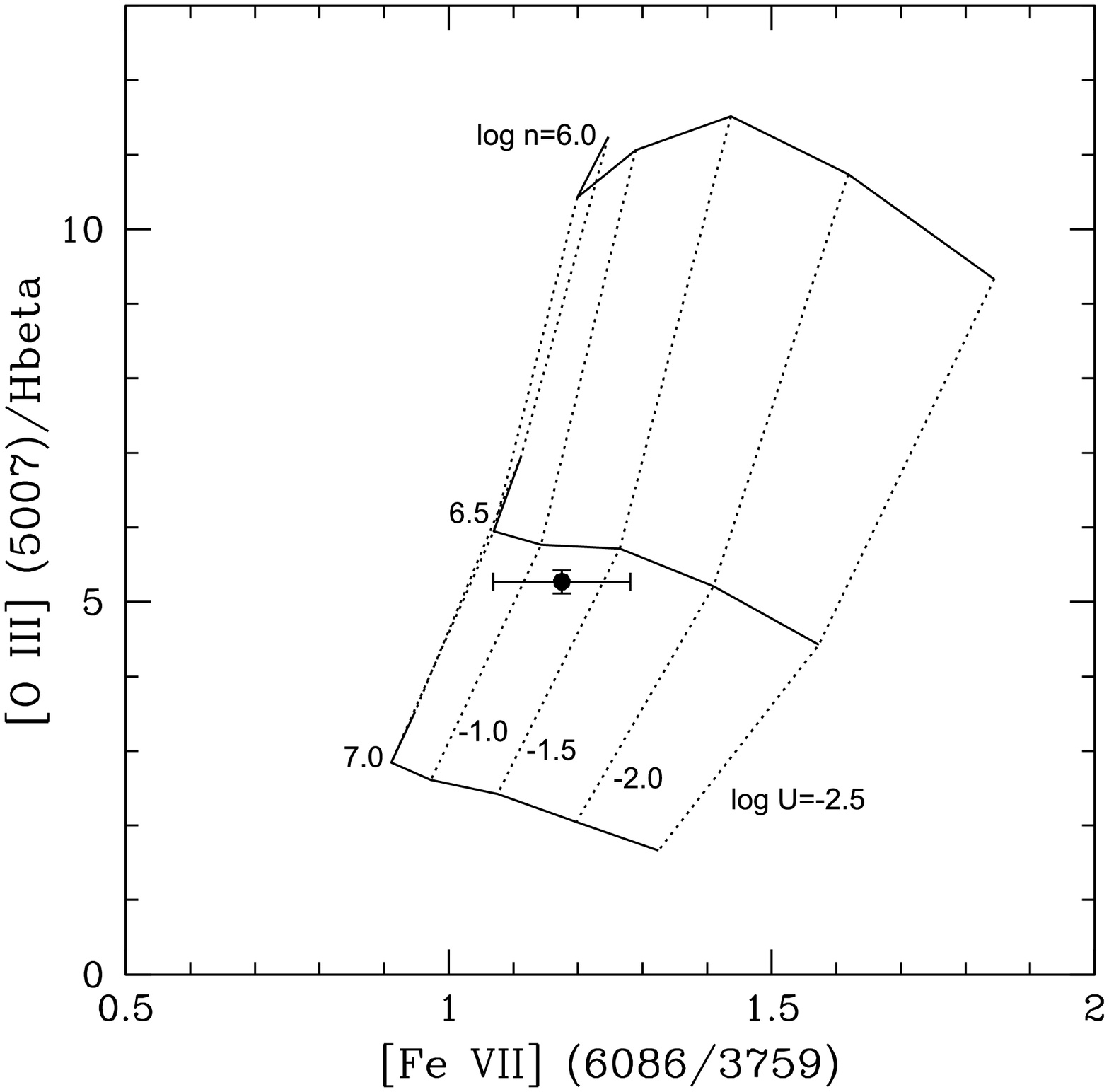}
\caption{Diagnostic plot showing the measured [O \textsc{iii}]$\lambda$5007/H$\beta$ and [Fe~\textsc{vii}](6086/3759) ratios (filled point), compared with photoionisation model results for different densities and ionisation parameters. Equal densities are connected by the solid lines (as labelled), and the ionisation parameters are connected by the dotted lines (also labelled).}
\end{figure}

\begin{figure}
\centering
\includegraphics[scale=0.38]{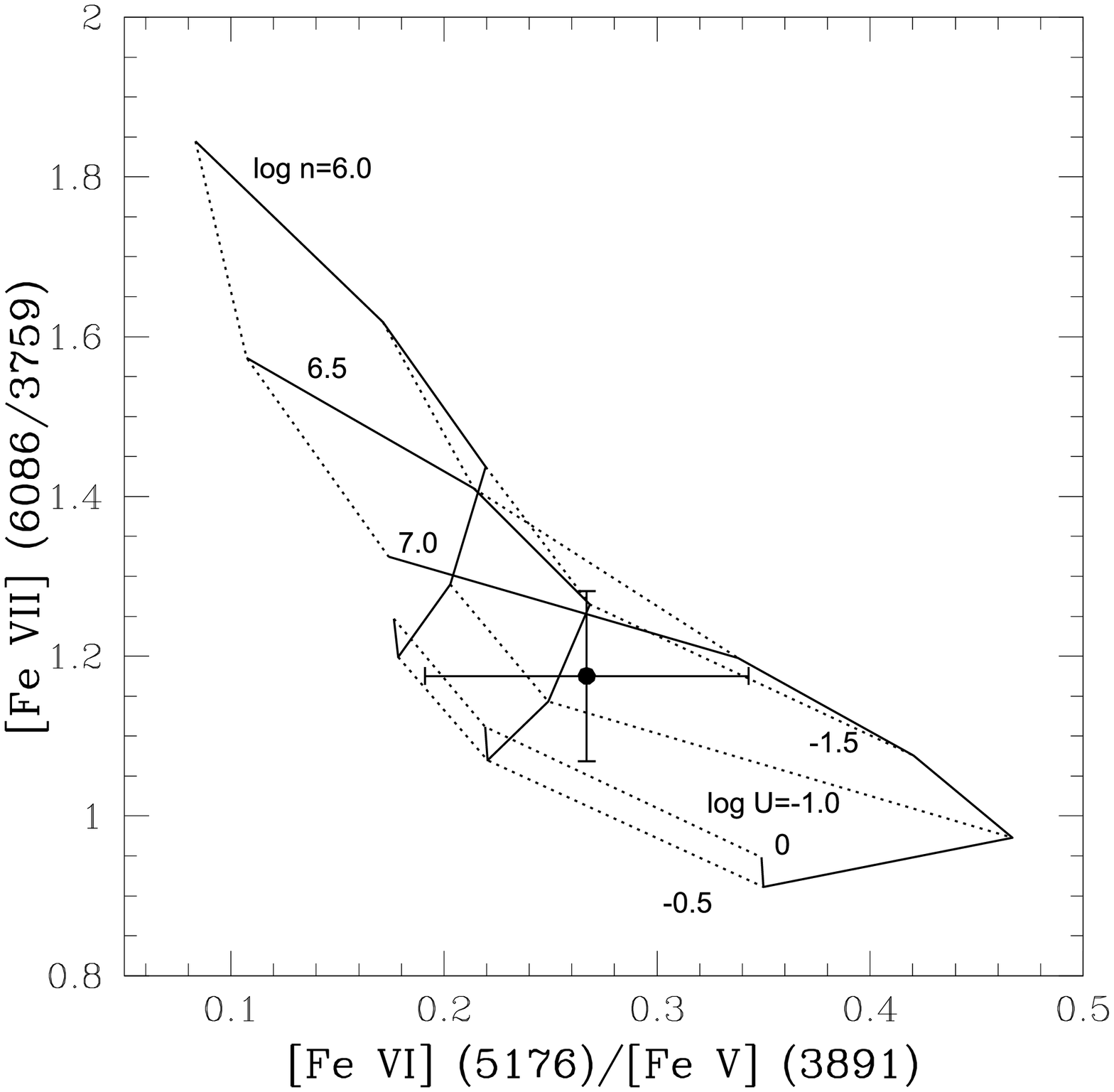}
\caption{Diagnostic plot showing the measured [Fe \textsc{vii}](6086/3759) and [Fe \textsc{vi}]$\lambda$5176/[Fe \textsc{v}]$\lambda$3891 ratios (filled point), compared with photoionisation model results for different densities and ionisation parameters. Equal densities are connected by the solid lines (as labelled), and the ionisation parameters are connected by the dotted lines (also labelled)..}
\end{figure}

\begin{figure}
\centering
\includegraphics[scale=0.38]{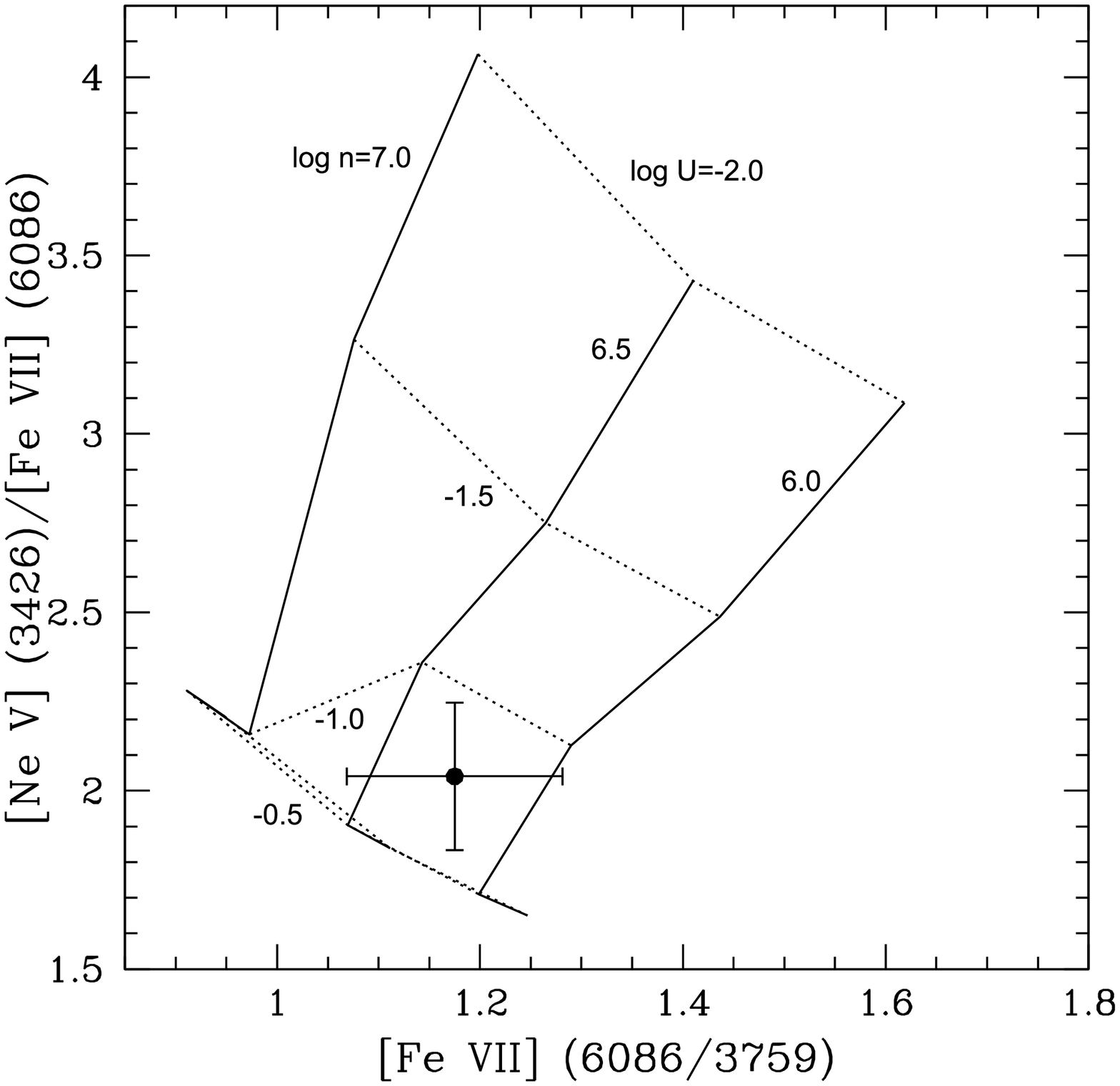}
\caption{Diagnostic plot showing the measured [Ne \textsc{v}]$\lambda$3426/H$\beta$ and [Fe~\textsc{vii}](6086/3759) ratios (filled point), compared with photoionisation model results for different densities and ionisation parameters. Equal densities are connected by the solid lines (as labelled), and the ionisation parameters are connected by the dotted lines (also labelled).}
\end{figure}

\begin{figure}
\centering
\includegraphics[scale=0.38]{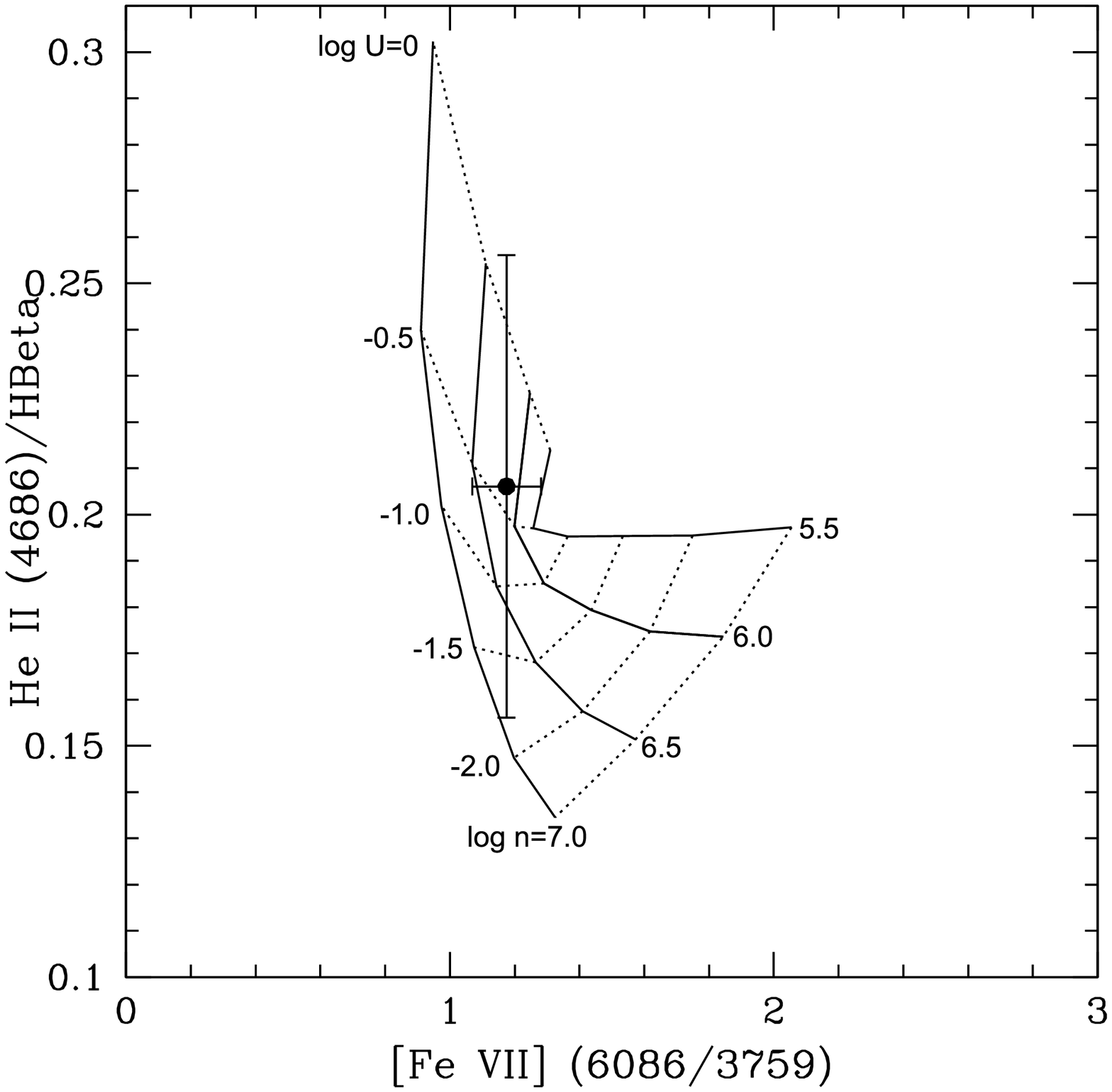}
\caption{Diagnostic plot showing the measured He \textsc{ii}$\lambda$4686/H$\beta$ and [Fe \textsc{vii}](6086/3759) ratios (filled point), compared with photoionisation model results for different densities and ionisation parameters. Equal densities are connected by the solid lines (as labelled), and the ionisation parameters are connected by the dotted lines (also labelled).}
\end{figure}

\begin{figure}
\centering
\includegraphics[scale=0.38]{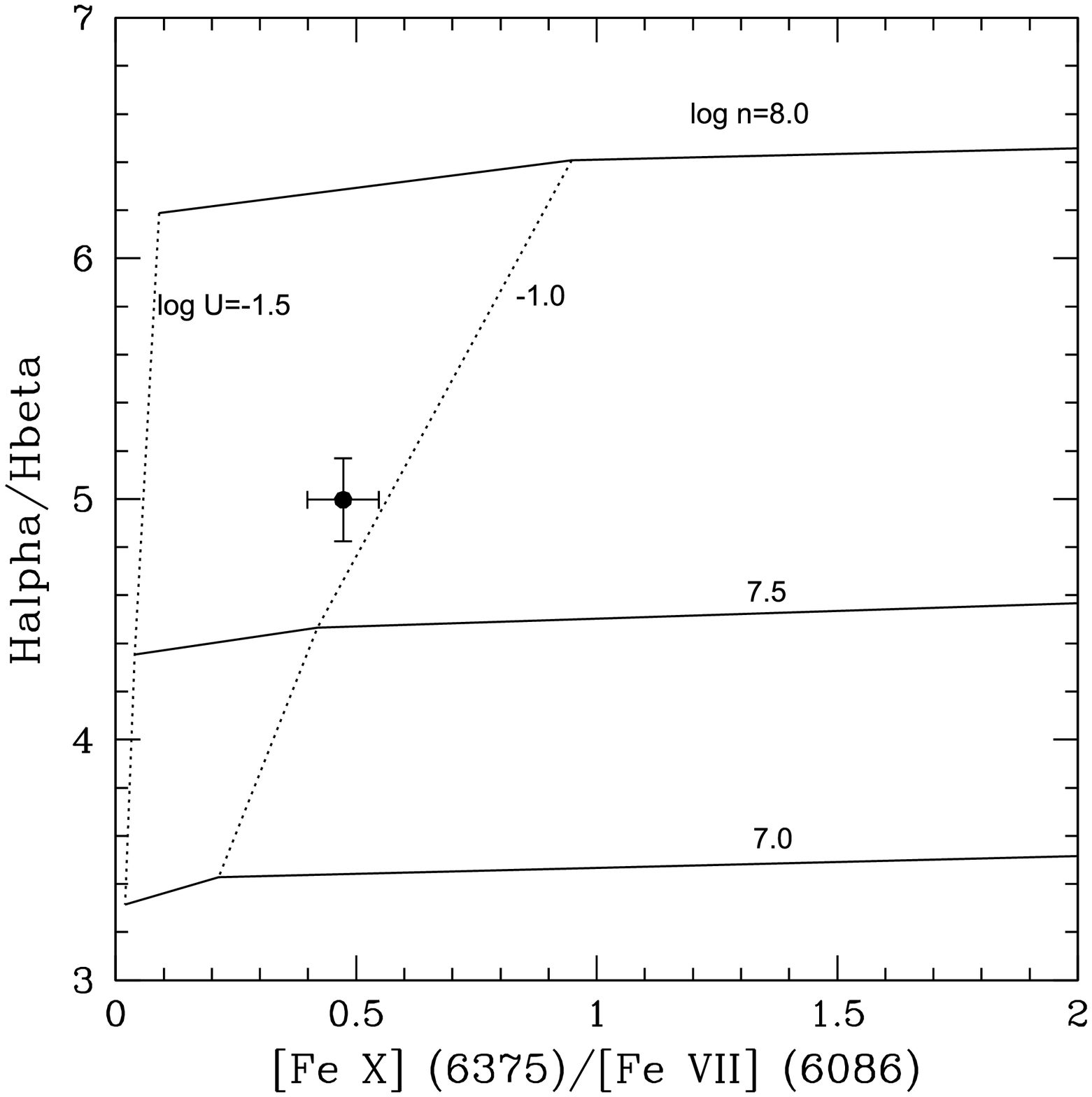}
\caption{Diagnostic plot showing the measured H$\alpha$/H$\beta$ and [Fe \textsc{x}]$\lambda$6374/[Fe \textsc{vii}]6086 ratios (filled point), compared with photoionisation model results for different densities and ionisation parameters. Equal densities are connected by the solid lines (as labelled), and the ionisation parameters are connected by the dotted lines (also labelled)..}
\end{figure}

To investigate the physical conditions implied by both the low ionisation and high ionisation species, the photoionisation model code CLOUDY\footnote{The version of CLOUDY used in this investigation is c08.00.} \citep{ferland} was used to create single slab photoionisation models for the emission lines of Q1131+16. The ionisation parameter was varied over the range -3.0 $\leq$ log[U] $\leq$ 0, in steps of log[U]=0.5, and the hydrogen density was varied over the range 3.0 $\leq$ log($n_H$ cm$^{-3}$) $\leq$ 8.0 in steps of log($n_H$ cm$^{-3}$) =0.5. The rest of the properties of the model were kept constant: an ionising power-law index of -1.2 was assumed\footnote{Testing power-laws between -1.5 and -0.8, this power-law was the most successful at modelling the observed He \textsc{ii}$\lambda$4686/H$\beta$ ratio compared to the FHIL ratios.}, and we only considered radiation-bounded models that are optically thick to the ionising continuum. In terms of abundances, Solar 84 was used (see \citealt{ga}, \citealt{gn}). Also no grains were included. This was so that the CLOUDY results were not affected by depletion of iron due to condensation onto dust grains. \citet{nagao3} have demonstrated for a large sample of AGN that the observed FHIL ratios\footnote{The ratio used in \citet{nagao3} was [Fe \textsc{vii}]$\lambda$6086/[Ne \textsc{v}]$\lambda$3426.} cannot be reproduced in CLOUDY models that contain grains, but can be successfully reproduced with models not containing grains. 

The ratios measured for each of the main diagnostic ratios used in this study are given in Table 3. The WHT ratios are used in this investigation because the WHT spectrum has the greatest wavelength range.

Considering first the low ionisation lines, we plot the transauroral [O\textsc{ii}](3726$+$3729)/(7317$+$7330) and [S\textsc{ii}](6717$+$6731)/(4069$+$4076) ratios against each other in Figure 14. By comparison with the models, the ratios are consistent with hydrogen densities in the range 3.5 $<$ log($n_H$ cm$^{-3}$) $<$ 4.5 at ionisation parameters -3.0 $<$ log[U] $<$ -2.0. The densities and ionisation parameters found for these low ionisation ratios are typical of these found for the NLR of AGN in general (e.g. see \citealt{peterson},  \citealt{osterbrock3}).

On the other hand, the high ionisation species show markedly different results. In Figure 15 the [Fe \textsc{vii}](6086/3759) ratio is plotted against the [O \textsc{iii}](5007/4363) ratio. These ratios consistent with hydrogen densities 6.0 $<$ log($n_H$ cm$^{-3}$) $<$ 6.5 at ionisation parameters in the range -1.5 $<$ log[U] $<$ 0. Similarly Figure 16 shows the results found by comparing the [Fe \textsc{vii}] ratios of (6086/3759) and (5159/6086). These ratios consistent with hydrogen densities 5.5 $<$ log($n_H$ cm$^{-3}$) $<$ 7.0 and ionisation parameters -1.0 $<$ log[U] $<$ 0. The densities and ionisation parameters indicated by the latter ratios are much higher than those implied by the low ionisation emission lines.

As mentioned previously, the [O \textsc{iii}]$\lambda$5007 emission line intensity is surprisingly low relative to H$\beta$ for an AGN with a high ionisation emission line spectrum. Figure 17 illustrates that such a low ratio is possible at high densities. The [O \textsc{iii}]$\lambda$5007/H$\beta$ emission ratio is consistent with hydrogen densities 6.5 $<$ log($n_H$ cm$^{-3}$) $<$ 7.0 at ionisation parameters in the range -1.5 $<$ log[U] $<$ 0. At such densities, above the critical densities of the [O \textsc{iii}]$\lambda\lambda$5007,4959 transitions, the [O \textsc{iii}] emission lines are suppressed relative to H$\beta$; this explains the relatively low [O \textsc{iii}]$\lambda$5007/H$\beta$ ratio.

FHILs with lower ionisation potentials, such as [Fe \textsc{vi}]$\lambda$5176, are also consistent with the results found using the [Fe \textsc{vii}] ratios. In Figure 18 the [Fe \textsc{vii}] (6086/3759) ratio is plotted against the [Fe \textsc{vi}]$\lambda$5176/[Fe \textsc{v}]$\lambda$3891 ratio. These ratios are consistent with hydrogen densities 6.5 $<$ log($n_H$ cm$^{-3}$) $<$ 7.0 at ionisation parameters in the range -1.5 $<$ log[U] $<$ 0. 

Similarly, FHILs which are not from iron ions are consistent with the results found using the [Fe \textsc{vii}] ratios. Figure 19 shows the [Ne \textsc{v}]$\lambda$3426/[Fe \textsc{vii}]$\lambda$6086 ratio plotted against the [Fe \textsc{vii}](6086/3759) ratio. This diagram is consistent with hydrogen densities 5.5 $<$ log($n_H$ cm$^{-3}$) $<$ 6.5 and ionisation parameters -1.0 $<$ log[U] $<$ -0.5. Therefore the high densities and ionisation parameters found for the FHILs are not limited to iron ions.

The He\textsc{ii}$\lambda$4686/H$\beta$ ratio is relatively insensitive to the density of the emission region, but more sensitive to the shape of the ionising continuum spectrum (e.g. \citealt{pf}), and whether the emitting regions are matter bounded \citep{rob}. Therefore, this ratio can be used to check whether the emission clouds are radiation bounded, and also whether the assumed ionising continuum shape is appropriate. The measured He\textsc{ii}$\lambda$4686/H$\beta$ ratio is found to be 0.21$\pm$0.05 and 0.23$\pm$0.01 for the WHT and Gemini spectra respectively. Figure 20 shows this ratio plotted against the [Fe \textsc{vii}](6086/3759) ratio. The densities and ionisation parameters implied by this diagram are consistent with those deduced from the other diagnostic diagrams involving the FHILs (5.0 $<$ log($n_H$ cm$^{-3}$) $<$ 7.0, -1.5 $<$ log[U] $<$ 0). In this case, the consistency between the data and the models, as well as the results for other high ionisation line ratios, support the assumption of optically thick models with an ionising continuum spectral index of -1.2.

Although the models can successfully reproduce the FHIL line ratios in Table 3, the emission line ratios involving lines with the highest critical densities (e.g. H$\alpha$/H$\beta$, [Fe \textsc{x}]/[Fe \textsc{vii}]) cannot be successfully modelled using the range of physical conditions and ionisation parameters that successfully reproduce the [Fe \textsc{vii}] and [O \textsc{iii}] line ratios. Indeed the higher critical density line ratios can only be reproduced at higher densities ($n_H$ $>$ 10$^7$ cm$^{-3}$) using CLOUDY. Figure 21 shows H$\alpha$/H$\beta$ ratio plotted against the [Fe \textsc{x}]$\lambda$6374/[Fe \textsc{vii}]$\lambda$6086 ratio. In this case, hydrogen densities of 7.5 $<$ log($n_H$ cm$^{-3}$) $<$ 8.0 and ionisation parameters -1.5 $<$ log[U] $<$ -1.0 are required to reproduce the data. This suggests that some regions in Q1131+16 have even higher densities than implied by the [Fe \textsc{vii}], [Fe \textsc{vi}] and [O \textsc{iii}] diagnostic ratios.

To summarise, most of the key high and low ionisation diagnostic ratios are consistent with the radiation-bounded photoionisation models. While the low ionisation lines require low densities (3.5 $<$ log($n_H$ cm$^{-3}$) $<$ 4.5) and ionisation parameters (-3.0 $<$ log[U] $<$ -2.0), the higher ionisation lines imply much higher densities (6.0 $<$ log($n_H$ cm$^{-3}$) $<$ 8.0) and ionisation parameters (-1.5 $<$ log[U] $<$ 0). However, it is important to emphasise that our approach assumes that the lines contributing to a particular ratio are emitted by a single cloud with a single density and ionisation parameter; the reality is likely to be more complex. In particular, some of the low ionisation lines (e.g. [S \textsc{ii}]$\lambda$4072, [O\textsc{ii}]$\lambda$7330) may have contributions from the high density/high ionisation regions, while some of the higher ionisation lines (e.g. [O \textsc{iii}]$\lambda$5007, H$\beta$) will have contributions from the low density/low ionisation region. Therefore the differences deduced between the conditions in the high and low density regions may be even more extreme than implied by our simple analysis. 

\subsection{Explaining the high H$\alpha$/H$\beta$ ratio}

The H$\alpha$/H$\beta$ ratio gives an unusual result in Q1131$+$16. This may be due to the high density of the FHIL region in this object coupled with an extended partially ionised zone \citep{gaskell}. Such high densities can lead to significant collisional excitation of the neutral hydrogen atoms in the partially ionised zones of clouds in the FHIL region. This results in an enhanced H$\alpha$ emission, but no further enhancement in the other Balmer series emission lines, which have lower cross-sections for collision, as well as higher excitation temperatures (\citealt{gaskell}, \citealt{osterbrock3}). Such a process could explain the high H$\alpha$/H$\beta$ ratio measured in the spectrum of Q1131$+$16, since there is likely to be be a major contribution from the FHIL region to the line flux of H$\alpha$. 

The H$\alpha$/H$\beta$ flux ratio was investigated using CLOUDY, as described in $\oint$ 4.1. The ratio indicated by the WHT data could only be reached at the highest densities ($n_H$ $>$ 10$^7$ cm$^{-3}$: see Figure 21) and is not consistent with the conditions implied by the [Fe \textsc{vii}] and lower iron species ratios. Note that high H$\alpha$/H$\beta$ would also be favoured by lower metal abundances and a harder ionising continuum than we have assumed in our modelling \citep{gaskell}.

\subsection{Location of the FHIL region}

The presence of FHILs in the spectra of the majority of active galaxies has led to a number of studies on their nature, kinematics and location in AGN (e.g. see \citealt{mullaney}). One of the most credible suggestions for the origin of the FHIL emission is the inner torus wall (see \citealt{murayama2} and \citealt{nagao1}). The FHIL emission is rich in iron lines, which can be enhanced (relative to the classical NLR) by the evaporation of dust grains in the inner torus wall, releasing the iron locked up in the grains \citep{pier}. Moreover, studies of molecular emission from the circum-nuclear molecular clouds of the centres of the Milky Way and other galaxies (hypothetically the torus) reveal densities in the range n${_{H2}}$ $\sim$ 10$^{5-6}$ cm$^{-3}$ (e.g. \citealt{pag}), similiar to the densities deduced for the FHILs. 

In the case of Q1131+16, our results are consistent with the idea that the FHILs are emitted by the $\it{far}$ $\it{wall}$ of the torus, observed with our line of sight at a relatively large angle to the  torus axis. We emphasise that such a geometry is consistent with the relatively narrow line widths and small velocity shifts of the FHIL, since the circular velocities associated with the torus and any out-of-the-plane gas motions would then be directed close to perpendicular to the line of sight. In this case, our results imply that the torus is dynamically cold, with velocity dispersion that is small relative to its circular velocity.

It is difficult to entirely rule out the idea that the FHILs are emitted by an infalling molecular cloud, or a clump of the torus which has broken off from the rest of the torus and is falling towards the AGN. However, in such cases we might expect to see a larger velocity shift between the FHIL emission lines and the host galaxy rest-frame, unless the cloud happens to be falling perpendicular to the line of sight.

Therefore the FHIL emission is most likely to originate from the inner torus wall of the AGN, supporting the original suggestion of \citet{murayama2} and \citet{nagao1}.

\subsection{The radial distance to the FHIL region}

The idea that the FHIL region is located in the torus can be further investigated by determining its radial distance from the AGN ($r_H$).

First we estimate $r_H$ based on the spatial information given in $\oint$3.1, and the parameters determined in $\oint$4.1. The ionising luminosity ($L_{ION}$) of the illuminating source is related to the radial distance of the emission region from the ionising source (r) and the ionisation parameter (U), by:
\begin{equation}
   L_{ION}=4\pi r^2Un_Hc\{ h\nu  \}_{ion};
\end{equation}
where $n_{H}$ is the hydrogen density, $\left \langle h\nu  \right \rangle_{ion}$ is the mean ionising photon energy\footnote{We assume a value of 56 eV for $\left \langle h\nu  \right \rangle_{ion}$, which is based on an ionising power-law of -1.2, with photon energy limits of 13.6 eV and 5 keV.}, and c is the speed of light. Then, solving for the radial distance of the low ionisation/density region, we get: 
\begin{equation}
r{_{L}}^{2}=\frac{L_{ION}}{4\pi\ U_{L}n^{L}_{H}c\left \langle h\nu  \right \rangle_{ion}}
\end{equation}
and similarly for the high ionisation/density emission region:
\begin{equation}
r{_{H}}^{2}=\frac{L_{ION}}{4\pi\ U_{H}n^{H}_{H}c\left \langle h\nu  \right \rangle_{ion}}.
\end{equation}
By combining equations 2 and 3, we obtain an equation which allows us to determine the relative radial distances of the high and low density regions from the AGN:
\begin{equation}
 \left [ \frac{r_{H}}{r{_{L}}}\right ]^{2}=\frac{U_{L}n^{L}_{H}}{U_{H}n^{H} _{H}}.
\end{equation}
Using the densities and ionisation parameters estimated in $\oint$4.1, we find that the radial distance of the FHIL emission region ranges from 0.0002 to 0.03 of that of the low ionisation region. As mentioned in $\oint$3.6, the [O \textsc{ii}] $\lambda$3727 emission is resolved in the 2D spectrum of Q1131+16, and the half width at half maximum (HWHM) measured for the central peak of this emission line is found to be 0.53$\pm$0.05 arcseconds (corrected for the seeing). This corresponds to $r_L$=1.7$\pm$0.2 kpc, implying that the radial distance of the FHIL region from the AGN is in the range 0.30 $<$ $r_H$ $<$ 50 pc --- consistent with recent observational estimates of the size of the torus in nearby AGN, based on infrared interferometry and high resolution molecular line observations (e.g. \citealt{haschick}, \citealt{chou}, \citealt{tristram}). However, we emphasise that this argument assumes that the [O \textsc{ii}] emission line region is solely ionised by the AGN, which may not be true if stellar photoionisation is significant at a kpc scale. Also, if the contrast between the conditions in the low and high ionisation regions is larger than assumed here (see discussion in $\oint$ 4.1), the radial distance of the FHIL region from the AGN will be correspondingly lower. Overall it is likely that this analysis, based on estimates of U and $n_H$, overestimates the true radial distance of the FHIL region.

If the ionising luminosity ($L_{ION}$) is known it is also possible to estimate the radial distance of the FHIL region from the AGN using equation 1 directly. $L_{ION}$ can be estimated from the bolometric luminosity ($L_{BOL}$), which itself can be determined from the [O \textsc{iii}]$\lambda$5007 emission line luminosity ($L_{[O\textsc{iii}]}$) by manipulating equation 1 from \citet{dicken}\footnote{Assuming $L_{[O\textsc{iii}]}$=5.5$L_{H\beta}$ rather than $L_{[O\textsc{iii}]}$=12$L_{H\beta}$.} to give:
 \begin{equation}
L_{BOL}=\frac{100}{C_{nlr}}L_{[O\textsc{iii}]},
\end{equation}
where $C_{nlr}$ is the NLR covering factor. Assuming typical NLR covering factors in the range 0.02 $<$ $C_{NLR}$ $<$ 0.08 \citep{nl}, we estimate a bolometric luminosity in the range 2x10$^{38}$ $<$ $L_{BOL}$ $<$ 10$^{38}$ W.

Alternatively, we can use our knowledge of the likely geometry of the FHIL to estimate an upper limit on the covering factor of the FHIL region and a lower limit on $L_{BOL}$. Assuming that the FHILs are emitted by the inner wall of the torus on the far side of the AGN (see $\oint$ 4.3), we see a maximum of 25\% of the total area of the inner wall of the torus (with the rest obscured). A typical torus with an opening angle of 45$^{\circ}$ has a $\it{total}$ covering factor of $C_{torus}$=0.7. Therefore, for our assumed geometry, an upper limit on the covering factor of the FHIL region is $C_{FHIL}$=0.7x0.25=0.175. Substituting this into equation 5, we obtain the following lower limit on the bolometric luminosity: $L_{BOL}$ $>$ 10$^{38}$ W.

We convert $L_{BOL}$ to $L_{ION}$ using the relationship given in \citealt{elvis2} ($L_{BOL}$ $\approx$ 3.1 $L_{ION}$), which leads to 3x10$^{37}$ $<$ $L_{ION}$ $<$ 3x10$^{38}$ W. Then, using equation 1 and the model parameters in $\oint$ 4.1, we estimate the radial distance of the FHIL region from the AGN to be in the range 0.3 $<$ $r_{H}$ $<$ 30 pc. For comparison, based on our estimates of the AGN bolometric luminosity, the sublimation radius of the dust grains in the torus\footnote{We calculate the sublimation radius using the expression for AGN sublimation radius given in \citealt{elitzur}.} is 0.26 $<$ $r_{sub}$ $<$ 0.52 pc --- consistent with the lower end of the range of our estimates of $r_{H}$. This consistency supports the idea that the FHILs are emitted by the inner torus wall.

\subsection{The unusual strength of the FHIL}

It is also important to consider why the FHILs are so much stronger in Q1131+16 and other similar objects than in more typical AGN.

First we must consider the possibility that, rather than an AGN, the spectrum of Q1131+16 is the result of a supernova, in particular a type IIn supernova. A type IIn supernova is a subclass of supernova for which it is believed that the supernova progenitor has undergone mass stripping of its outer layers in the late stages of its life. This process introduces a dense hydrogen gas in the circumstellar medium surrounding the progenitor. The shockwave from the supernova interacts with the material, heating it. This results in a rich spectrum of narrow emission lines, which has some similar characteristics of the spectrum of Q1131+16 \citep{smith}. It is notable that early observations of the [Fe \textsc{vii}] emission lines in one such supernovae, SN 2005ip (see \citealt{smith}), indicate electron densities of 10$^8$ cm$^{-3}$ and higher \citep{smith}.   

One argument against this suggestion is that the [O \textsc{iii}]$\lambda$5007 luminosity of Q1131+16 exceeds that of the hottest known type IIn supernova by 3 orders of magnitude \citep{smith}. Also the line emission from a supernova is expected to be variable. In the case of the type IIn supernova considered by \citet{smith}, the [O \textsc{iii}]$\lambda$5007 increased 8-fold after 100 days and then faded to half this intensity after 3 years. The [Fe \textsc{vii}]$\lambda$5720 emission line decreased 4-fold in intensity after 100 days, but after 3 years this intensity had increased to half its original intensity. As for the [Fe \textsc{vii}]$\lambda$6086 emission line, this decreased in intensity by half after 100 days and then varied over 3 years around half its original intensity. However, in Q1131+16 there is no evidence for emission lines fading in the 3 years between the WHT and Gemini spectra. This lack of variability rules out any supernova component.

The lack of variability in the emission lines of Q1131+16, also provides strong evidence against the the idea that its unusual FHILs are due to an illumination flare caused by the tidal disruption of a star by the SMBH \citep{komossa1b}. In the well-studied case of SDSSJ095209.56+214313.3 \citep{komossa2}, FHILs such as [Fe \textsc{vii}]$\lambda\lambda$3759\&5159, are seen to dramatically vary over a timescale of 3 years: the [Fe \textsc{vii}]$\lambda$3759 line\footnote{Here all ratios are relative to [O \textsc{iii}]$\lambda$5007. In SDSSJ095209.56+214313.3 the [O \textsc{iii}] flux varies very little between the spectra (see \citealt{komossa2}).} by 36\% in intensity over 3 years, the [Fe \textsc{vii}]$\lambda$5159 line by 79\%, and He\textsc{ii}$\lambda$4686 by 65\%. In contrast, the same lines in Q1131+16 did not vary significant over a similar 3 year timescale. Moreover, the Balmer lines in Q1131+16 lack the variable, multi-peaked profiles observed in SDSSJ095209.56+214313.3 (\citealt{komossa1b} \& \citealt{komossa2}). 

\begin{figure*}
\centering
\subfloat[Line of sight]{\includegraphics[scale=0.3]{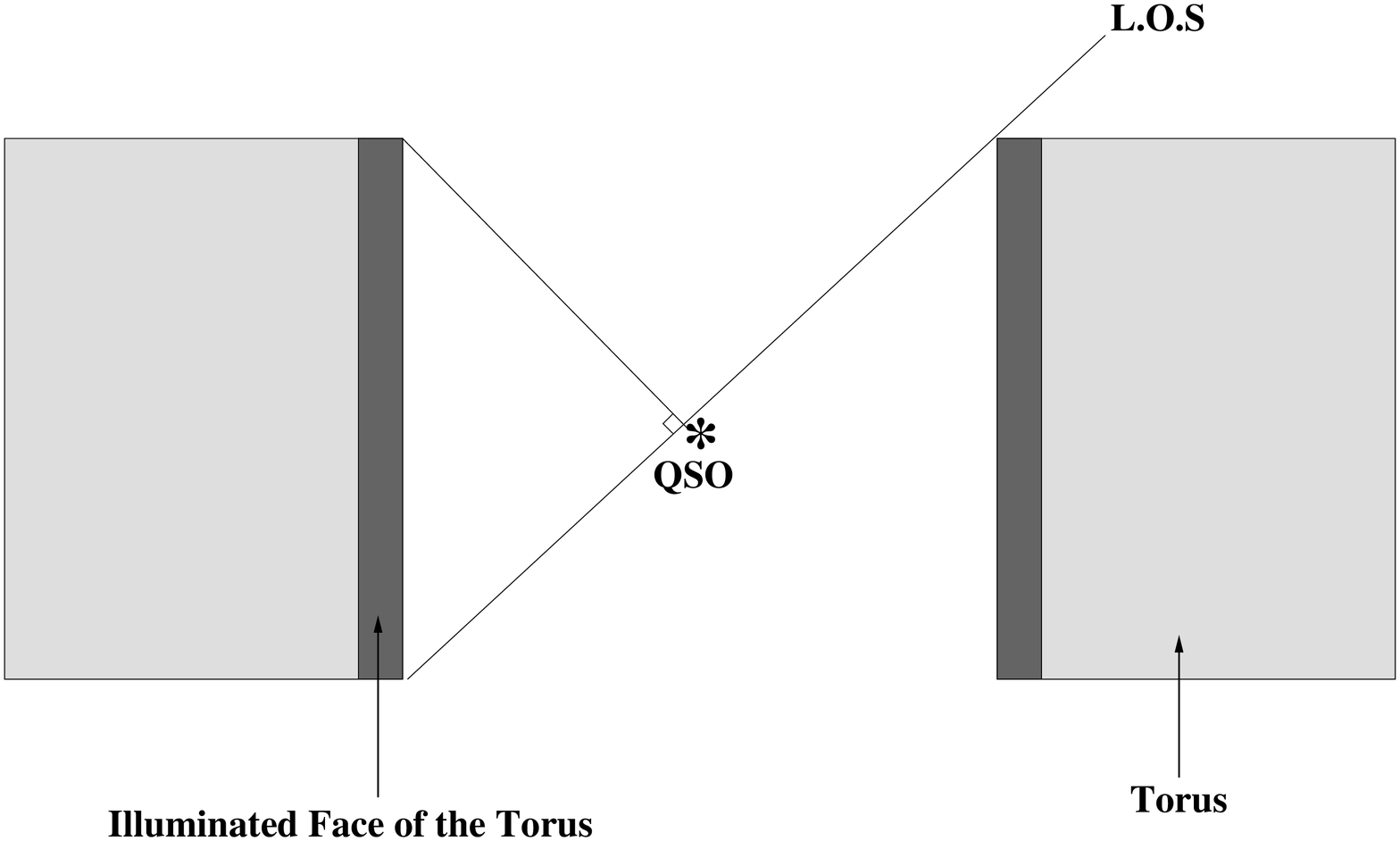}}\hspace{40pt}
\subfloat[Torus surface area]{\includegraphics[scale=0.33]{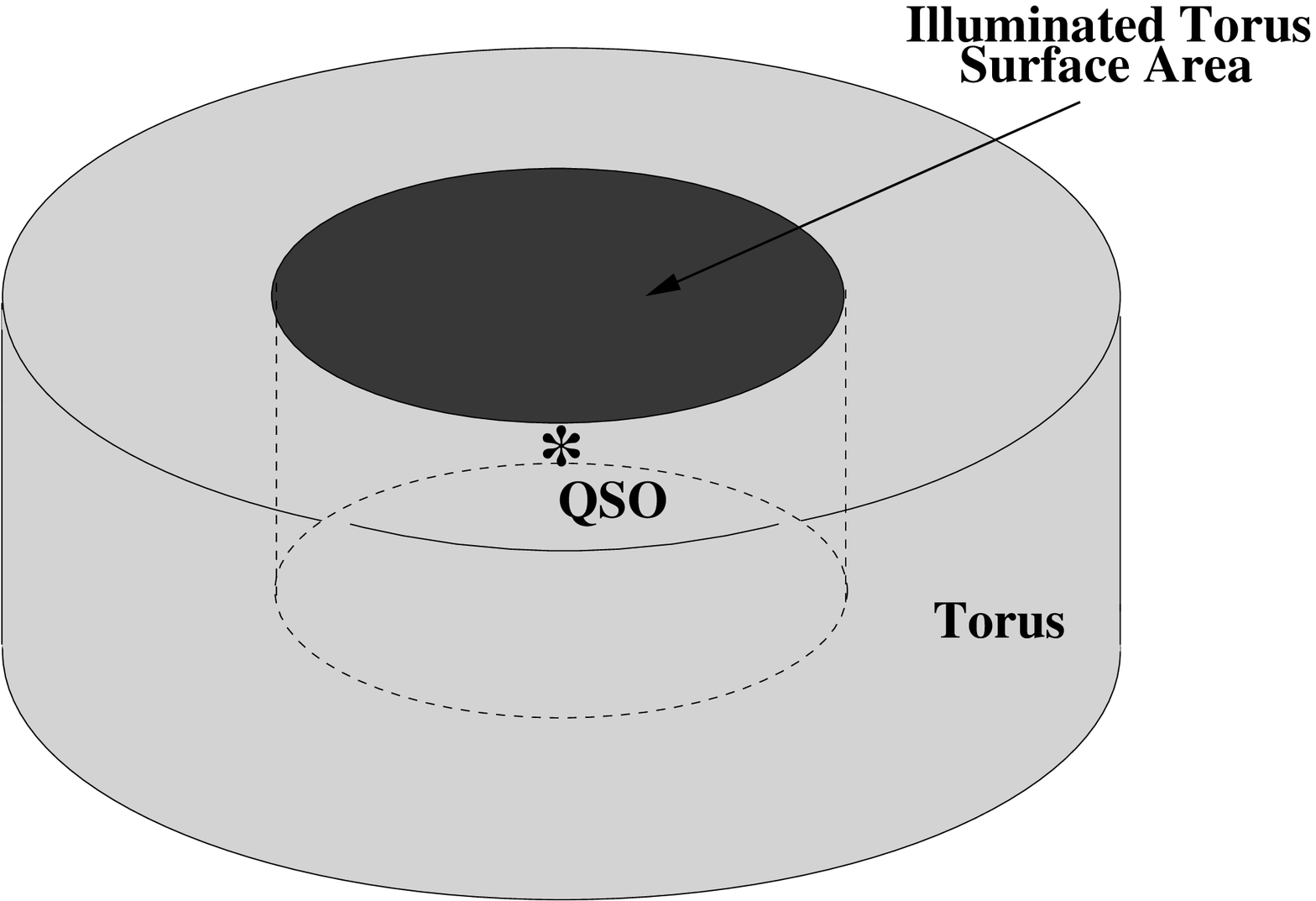}}
\caption{\textit{(a)} Schematic displaying an edge-on view of the possible observed orientation of the AGN of Q1131+16. \textit{(b)} Schematic displaying the visible area of the inner torus wall of Q1131+16 as viewed by the observer. This special geometry will maximise the observed emission line flux of the inner wall of the torus yet still obscure the central quasar, thus lowering the underlying continuum, and hence resulting in FHILs with large equivalent widths.}
\end{figure*}

Rather than an illumination event, an explanation for the unusual strength of the FHILs may by provided by the inner torus wall being viewed at a specific orientation, at which the inner wall on the far side of the torus can be seen by the observer, but the quasar itself remains hidden, because the torus is being viewed at a relatively large angle to its axis (see $\oint$ 4.3). In this case, the visible area of the torus wall is high enough that the FHILs are strong, but the direct quasar continuum emission does not dilute the torus line emission (see Figure 22). Therefore the FHILs are observed with large equivalent widths. Viewed at smaller angles to the axis of the torus, the luminous quasar nuclear emission would become directly visible, substantially reducing the equivalent widths of the FHILs. 

In this context we note that, despite the unusually large equivalent widths of its FHILs, the measured ratios of the FHILs to  lower ionization lines (e.g. [Fe \textsc{vii}]/[O \textsc{iii}], [Fe \textsc{vii}]/[S \textsc{ii}], [Fe \textsc{vii}]/[O \textsc{i}] and [Fe \textsc{x}]/[O \textsc{i}]) for Q1131+16 fall well within the ranges measured for typical broad line AGN (Sy1, Sy1.5, NLSy1) in the sample compiled by \citet{nagao2}. Therefore, viewed from a direction closer to the torus axis, we would expect Q1131+16 to display a normal quasar/Seyfert 1 spectrum, but the multitude of fainter FHIL would be difficult to detect because of the dilution by the bright quasar continuum and broad line emission.

\begin{table*}
\caption{Table comparing key properties determined for the FHIL regions in of Q1131+16, III Zw 77 \citep{osterbrock2}, Tololo 0109-383 (\citealt{fosbury}, \citealt{murayama2}) and ESO 138 G1 \citep{alloin}. *shows systematic increase from low to high ionisation lines.}
\begin{tabular}{r c c l}
\hline
Object & Density & Velocity Width & [O\textsc{iii}](5007/4363)\\
 & (cm$^{-3}$) & (FWHM, km s$^{-1}$) & \\
\hline
Q1131+16 & 10$^{8}$$>$$n_H$$>$10$^{5}$ & 361$\pm$29 & 5.86$\pm$0.21 (WHT)\\
III Zw 77 & $n_e$$>$10$^{5}$ & $\sim$400* & 11.48$\pm$0.02\\
Tololo 0109-383 & $n_e$$>$10$^{6}$ & $\sim$330 & 21.50$\pm$0.46\\
ESO 138 G1 & $n_e$$>$10$^{5}$ & $\sim$1000 & 30$\pm$2\\
\hline
\end{tabular}
\end{table*}

However, it is more challenging to  explain why the FHIL are significantly stronger relative to the low ionization lines in Q1131+16 than in typical Type 2 AGN \citep{nagao2}. Given the geometry in Figure 22, we would expect to observe significant FHIL
emission over a range of viewing angles, even if the quasar nucleus itself is obscured; rather
than a sharp cut-off, for this simple geometry, the strength of the FHIL would 
gradually diminish as the orientation of the axis of the torus to the
line of sight increased (i.e. the torus became more edge-on). Therefore, although the 
particular geometry shown in Figure 22 would maximise the FHIL emission, it is unlikely that a 
specific geometry alone can explain the rarity of
objects like Q1131+16. We now discuss three further factors that might lead
to the FHIL appearing relatively strong compared with the general population of
Type 2 AGN.
\begin{itemize}
\item [-] {\bf A lack of conventional NLR emission.} If the emission from a ``conventional'' NLR 
on a scale of $\sim$100~pc --- 1~kpc were strong, this would dilute the FHIL emission, and the overall
spectrum would not appear unusual relative to other Type 2 AGN. Therefore, the conventional NLR emission must be relatively weak in the case of Q1131+16 and other similar objects. Weak NLR emission can arise if the covering factor of the NLR is relatively small, or if the illuminating AGN has
only recently been triggered (within the last 100 -- 1000~yr), so that the ionising radiation has not
had time to traverse the scale of the NLR. However, while the lack of a conventional NLR might help to explain the unusual emission line ratios, it would not explain the unusually high equivalent
widths measured for many of the emission lines in Q1131+16. Indeed, if this was the dominant factor
we would expect the emission lines in Q1131+16 to have {\it lower} equivalent widths than most other Type 2 AGN, which is clearly not the case. 
\item [-] {\bf A lack of obscuration by large-scale dust structures that are not coplanar with the torus.}
Along with a compact torus on a scale of ~1 -- 100~pc, many active galaxies show evidence for larger
dust lane structures on a scale of ~100~pc -- 5~kpc. If the dust lanes are 
not exactly co-planar with the torus structures (e.g. thicker or different orientation), they
will make it more difficult to observe the inner face of the torus, thus reducing the strength
of the optical FHIL emission, and narrowing the range of angles over which
the FHIL would appear strong. While the lack such large-scale dust structures
would certainly increase the chance of observing unusually strong FHIL, their presence 
would in any case reduce the range of viewing angles over which
the FHIL would be visible.
\item [-] {\bf A recently triggered AGN.} If the AGN has only recently been triggered then, not only
might this result in reduced emission from the conventional NLR (see above), but the inner wall of the torus itself might not yet have reached an equilibrium radius set by dust sublimation. In this case, the clouds close to the AGN are likely to be in the process of evaporation by the
strong AGN radiation field \citep{pier} and, depending on
the density gradient, might have a higher ionisation parameter than clouds at the sublimation
radius, leading to relatively strong FHIL. The problem with this explanation for Q1131+16 is that
the radius of the FHIL emission line region is similar to the estimated sublimation radius of
the torus (see $\oint$ 4.4), suggesting an equilibrium situation. However, it is possible that future refinement of the physical conditions of the FHIL in Q1131+16 and the bolometric luminosity of its AGN might lead to a revision of this picture.
\end{itemize}
Overall, while a specific viewing angle is likely to provide an important part of the explanation for
the unusual spectrum of Q1131+16, one or more of the other factors discussed above may also be
significant. 

\subsection{Q1131+16 in relation to other similar objects}

As mentioned previously, there are a few similar objects in which the FHILs are unusually strong (III Zw 77, Tololo 0109-383 and ESO 138 G1). The key properties of these objects, as derived from their FHILs, are presented in Table 4. In particular, we note that the densities derived from the FHILs are consistently high. Also all the objects show a particularly low [O III](5007/4363) line ratio when compared to `typical' Seyfert galaxies (although in this respect Q1131+16 is the most extreme). Finally, all the objects show relatively modest FHIL line widths (FWHM), with the exception of ESO 138 G1 whose line widths are larger.

Because there are relatively few objects with such unusual features, it is entirely plausible that the properties of all these objects can be explained in a similar way e.g. with a  specific viewing angle of the torus that maximises the observed emission area of the torus whilst still obscuring the central engine, consistent with the scheme described by \citet{murayama2}, \citet{nagao1} and \citet{nagao2}.

\section{Conclusions}

Q1131+16 has a remarkable spectrum which displays a multitude of FHILs of relatively high equivalent width, as well as the more common emission lines expected of an AGN spectrum.

An in-depth study of the spectrum of Q1131+16 has revealed:

\begin{itemize}

\item emission line ratios for the FHIL region that are consistent with high densities (10$^{5.5}$ $<$ $n_H$ $<$ 10$^{8.0}$ cm$\textsuperscript{-3}$) and ionisation parameters (-1.5 $<$ log[U] $<$ 0);

\item similar velocity widths (FWHM) for the low and high ionisation lines, and a lack of evidence for substantial velocity shifts of the FHILs relative to the galaxy rest frame;

\item a small radial distance from the AGN for the region emitting the FHILs (0.30 $<$ $r_{FHIL}$ $<$ 50 pc).

\end{itemize}

Based on the high densities and relatively quiescent kinematics implied by our observations, it is likely that the FHILs in Q1131+16 are emitted by the torus wall, with the inner wall on the far side of the torus viewed directly by the observer, but the quasar itself remaining hidden. This geometry is also consistent with the relatively small radial distance found for the FHIL region.   

These results demonstrate the potential of the FHILs for probing the circum-nuclear obscuring regions in AGN. 

\section*{Acknowledgements}

M.R. acknowledges support in the form of an STFC PhD studentship. C.R.A. ackowledges financial support from STFC PDRA (ST/G001758/1). 
The authors acknowledge Jose Antonio Acosta Pulido for his valuable help
with the LIRIS data reduction. We would like to thank the referee for useful comments and suggestions. The authors acknowledge the data analysis facilities provided by the
Starlink Project, which is run by CCLRC on behalf of PPARC. The William Herschel Telescope and its service programme are operated on the
island of La Palma by the Isaac Newton Group in the Spanish Observatorio del
Roque de los Muchachos of the Instituto de Astrof\'{i}sica de Canarias.
This work is based on observations obtained at the Gemini Observatory, which
is operated by the
Association of Universities for Research in Astronomy, Inc., under a
cooperative agreement
with the NSF on behalf of the Gemini partnership: the National Science
Foundation (United
States), the Science and Technology Facilities Council (United Kingdom), the
National Research Council (Canada), CONICYT (Chile), the Australian Research
Council
(Australia), Minist\'{e}rio da Ci\^{e}ncia e Tecnologia (Brazil), and
Ministerio de Ciencia, Tecnolog\' ia e Innovaci\'on Productiva (Argentina).
The Gemini program under which these data were obtained is GS-2009B-Q-87.
 This research has made use of the NASA/IPAC Extragalactic Database (NED) which is operated by the Jet Propulsion Laboratory, California Institute of Technology, under contract with the National Aeronautics and Space Administration.

\label{lastpage}

\end{document}